\documentclass[useAMS]{mn2e}
\usepackage{times,psfig,epsfig}

\title[The baryon fraction in simulations of clusters]
{The baryon fraction in hydrodynamical simulations of galaxy clusters}

\author[S. Ettori et al.]
{S. Ettori$^1$, K. Dolag$^2$, S. Borgani$^{3,4,5}$, G. Murante$^6$ \\~\\
\footnotesize 
$^1$ INAF, Osservatorio Astronomico di Bologna, via Ranzani
  1, I-40127 Bologna, Italy (stefano.ettori@bo.astro.it) \\
$^2$ Max-Planck-Institut f\"ur Astrophysik, Karl-Schwarzschild Strasse
  1, Garching bei M\"unchen, Germany (kdolag@mpa-garching.mpg.de)\\
$^3$ Dipartimento di Astronomia dell'Universit\`a di Trieste, via
  Tiepolo 11, I-34131 Trieste, Italy (borgani@ts.astro.it)\\
$^4$ INFN -- National Institute for Nuclear Physics, Trieste,
  Italy\\ 
$^5$ CISC - Inderdept. Centre for Computational Sciences, University of 
  Trieste, Italy\\
$^6$ INAF, Osservatorio Astronomico di Torino, Strada Osservatorio 20,
  I-10025 Pino Torinese, Italy (murante@to.astro.it)
}

\begin{document}
\maketitle 

\begin{abstract}
We study the baryon mass fraction in a set of hydrodynamical simulations
of galaxy clusters performed using the Tree+SPH code {\tt GADGET-2}.
We investigate the dependence of the baryon fraction 
upon the radiative cooling, star formation,
feedback through galactic winds, conduction and redshift. 
Both the cold stellar component and the hot X-ray emitting gas have
narrow distributions that, at large cluster-centric distances $r \ga R_{500}$,
are nearly independent of the physics included in the simulations.
Only the non-radiative runs reproduce the gas fraction
inferred from observations of the inner regions ($r \approx R_{2500}$) of
massive clusters.
When cooling is turned on, the excess star formation 
is mitigated by the action of galactic winds,
but yet not by the amount
required by observational data. 
The baryon fraction within a fixed overdensity increases slightly
  with redshift, independent of the physical
  processes involved in the accumulation of baryons in the cluster
  potential well.
In runs with cooling and feedback, the increase in baryons is
  associated with a larger stellar mass fraction that arises at high
  redshift as a consequence of more efficient gas cooling.
For the same reason, the gas fraction appears less concentrated
at higher redshift.
We discuss the possible cosmological implications of our results and
find that two assumptions generally adopted, 
(1) mean value of $Y_{\rm b} = \frac{f_{\rm b}}{\Omega_{\rm b}/
\Omega_{\rm m}}$ not evolving with redshift, and (2) a fixed ratio between 
$f_{\rm star}$ and $f_{\rm gas}$ independent of 
radius and redshift, might not be valid. In the estimate of the 
cosmic matter density parameter, this implies some systematic
effects of the order of $\Delta \Omega_{\rm m}/\Omega_{\rm m} \la +0.15$ 
for non--radiative runs and $\Delta \Omega_{\rm m}/\Omega_{\rm m}$
$\approx +0.05$ and $\la -0.05$ for radiative simulations. 
\end{abstract} 
 

\section{Introduction}

Baryons in galaxy clusters, mainly in the form of stars in galaxies
and hot X-ray emitting plasma, trace the potential of the collapsed
structure in which they fall into. At the same time, their spatial
distribution and thermodynamical properties are affected by the
physical processes acting on them. 
Measurement of direct cluster observables, such as stellar
  optical light and hot gas X-ray emission, is the only viable approach to
  investigate the physics that drives the evolution of such structures.
The observed relations among proxies of
the internal energy and entropy levels show that, at least in massive
systems, the dominant baryonic component of the intra--cluster medium
(ICM) has a luminosity, temperature and mass that well follow those
predicted under the assumptions of a plasma emitting by bremsstrahlung
and sitting in hydrostatic equilibrium with the underlying dark matter
potential (e.g. Kaiser 1991, Evrard \& Henry 1991).

Moreover, the fact that galaxy clusters are relatively well isolated 
structures, that form in correspondence of the highest peaks
of the primordial gravitational fluctuations,
suggests that their relative baryon budget and the mass function are
highly sensitive tests of the geometry and matter content of the
Universe.  In particular, using X-ray observations of the baryonic
content to infer the gas and total mass of relaxed clusters in
hydrostatic equilibrium allows one to place a lower limit to the
cluster baryon mass fraction, which is expected to match the cosmic
value $\Omega_{\rm b}/\Omega_{\rm m}$ (White et al. 1993, Evrard 1997,
Mohr, Mathiesen \& Evrard 1999, Ettori \& Fabian 1999, Roussel, Sadat
\& Blanchard 2000, Allen et al. 2002, Ettori et al. 2003, Ettori 2003,
Allen et al. 2004).

The purpose of the present work is to use an extended set of
hydrodynamical simulations of galaxy clusters, treating a variety of
physical processes, to study how the spatial distribution of the
baryons, as contributed both by the stellar component and by the hot
X-ray emitting gas, are affected by the physical conditions within
clusters.
In this respect, our analysis extends
previous analyses of the baryon fraction in cluster simulations which
included only non--radiative physics (Evrard 1990, Metzler \& Evrard
1994, Navarro et al. 1995, Lubin et al. 1996, Eke et al. 1998, Frenk
et al. 1999, Mohr et al. 1999, Bialek et al. 2001), and the processes
of cooling and star formation (Muanwong et al. 2002, Kay et al. 2004,
Ettori et al. 2004, Kravtsov et al. 2005).
  
The paper is organized as follows: in Section~2, we describe our
dataset of simulated galaxy clusters; in Section~3, we present the
results obtained from this set of simulations on the gas and baryon
fraction, also comparing these results to the observational
constraints. In Section~4 we compare our results to previous analyses
based on different simulations. Finally, we summarize and discuss our
findings in Section~5.

\section{Properties of the simulated clusters}

We consider two sets of clusters, which have been selected from
different parent cosmological boxes. The first set is extracted from the
large--scale cosmological simulation presented in Borgani et
al. (2004).  The second one is a re-simulation of 9 galaxy clusters,
extracted from a pre-existing lower--resolution DM--only simulation. 
Our simulations were carried out with {\tt GADGET-2} (Springel
2005), a new version of the parallel Tree+SPH simulation code {\tt
GADGET} (Springel et al. 2001). It includes an entropy-conserving
formulation of SPH (Springel \& Hernquist 2002), radiative cooling, 
heating by a UV background, and a treatment of star formation and 
feedback from galactic winds powered by supernova explosions 
(Springel \& Hernquist 2003).

\subsection{Clusters extracted from a cosmological Box}

The first set of simulated clusters has been extracted from the
large-scale cosmological simulation of a ``concordance''
$\Lambda$CDM model ($\Omega_{0m}=0.3$, $\Omega_{0\Lambda}=0.7$,
$\Omega_{0\rm b}=0.019 h^{-2}$, $h=0.7$, $\sigma_8=0.8$; Borgani et
al. 2004).  Here we give only a short summary of its characteristics,
and refer to that paper for more details. The run follows the
evolution of $480^3$ dark matter particles and an equal number of gas
particles in a periodic cube of size $192 h^{-1}$ Mpc. The mass of the
gas particles is $m_{\rm gas}=6.89 \times 10^8 h^{-1} M_\odot$, while
the the Plummer-equivalent force softening is set to $7.5 h^{-1}$ kpc
from $z=0$ to $z=2$, while kept fixed in comoving units at higher
redshift. 
At $z=0$ we extract a set of 439 mass-selected clusters with virial
masses $M_{\rm vir}> 5\times10^{13}h^{-1} M_{\odot}$.

The efficiency of conversion of the energy provided by SN explosions
into a kinetic feedback (i.e. winds) is set to 50 per cent, which gives a
wind speed of $\approx$340 km s$^{-1}$.

\subsection{Resimulated clusters}

This set includes simulations of 
4 high mass systems with 
$M_{200} = $(1.1--1.8)$\times 10^{15} h^{-1} M_{\odot}$. The
cluster regions were extracted from a dark-matter only simulation with
a box-size of $479\,h^{-1}\,{\rm Mpc}$ of the same cosmological model
as the first set, but with a higher normalization of the power
spectrum, $\sigma_8=0.9$ (see Yoshida, Sheth \& Diaferio 2001).  Using
the ``Zoomed Initial Conditions'' (ZIC) technique (Tormen 1997), they
were re-simulated with higher mass and force resolution by populating
their Lagrangian volumes in the initial domain with more particles,
while appropriately adding additional high--frequency modes. The
initial particle distributions (before displacement) are of glass type
(White 1996).
%
The mass resolution of the gas particles in these simulations $m_{\rm
gas}=1.7\times 10^8\,h^{-1}\,M_\odot$. Thus, the clusters were
resolved with between $2\times10^6$ and $4\times10^6$ particles,
depending on their final mass.  For all simulations, the gravitational
softening length was kept fixed at
$\epsilon=30.0\,h^{-1}\,\mathrm{kpc}$ comoving (Plummer-equivalent),
and was switched to a physical softening length of
$\epsilon=5.0\,h^{-1}\,\mathrm{kpc}$ at $z=5$.

For simulations including star formation and feedback the SN
efficiency in powering galactic winds is set to 50 per cent,
as in the cosmological box, which turns into a wind
speed of $\approx$340 km s$^{-1}$. For the sake of comparison, some
runs have also been performed also by switched off the winds
completely or increasing the SN efficiency to unity.

For some of the cluster simulations we also included the effect of
heat conduction. Its implementation in SPH, which is both stable and
manifestly conserves thermal energy even when individual and adaptive
time--steps, has been described by Jubelgas, Springel \& Dolag (2004).
This implementation assumes an isotropic effective conductivity
parameterized as a fixed fraction of the Spitzer rate, that we assume
to be 1/3. It also accounts for saturation, which can become relevant
in low-density gas. For more details on the properties of simulated
galaxy clusters including thermal conduction see Dolag et al. (2004).

Furthermore, some simulations were carried out using a modified
artificial viscosity scheme suggested by Morris \& Monaghan (1997),
where every particle evolves its own viscosity parameter. Whereas in
this scheme shocks are as well captured as in the standard one,
regions with no shocks do not suffer for a residual non--vanishing
artificial viscosity. Therefore turbulence driven by fluid
instabilities can be much better resolved and as a result of this,
galaxy clusters simulated with this new scheme can build up sufficient
level of turbulence--powered instabilities along the surfaces of the
large scale velocity structure present in cosmological structure
formation (Dolag et al. 2005).

\begin{figure}
  \epsfig{figure=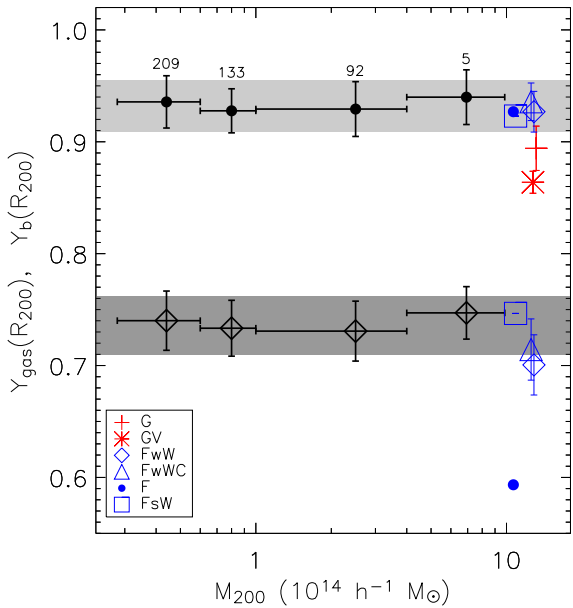,width=0.5\textwidth}
\caption{The baryon and gas mass fractions at $R_{200}$ and $z=0$ as
functions of the virial mass. The solid dots with error-bars refer to
the mean and dispersion measured in the corresponding bins in virial
mass.  The number of objects in each bin is indicated.  The most
massive re-simulated systems are also shown. 
The shaded regions indicate the $1 \sigma$ range of $Y_{\rm gas}$ 
and $Y_{\rm b}$ for the set of simulated clusters extracted 
from the cosmological box.
}\label{fig:ybar_m}
\end{figure}

\begin{figure}
\vbox{
 \epsfig{figure=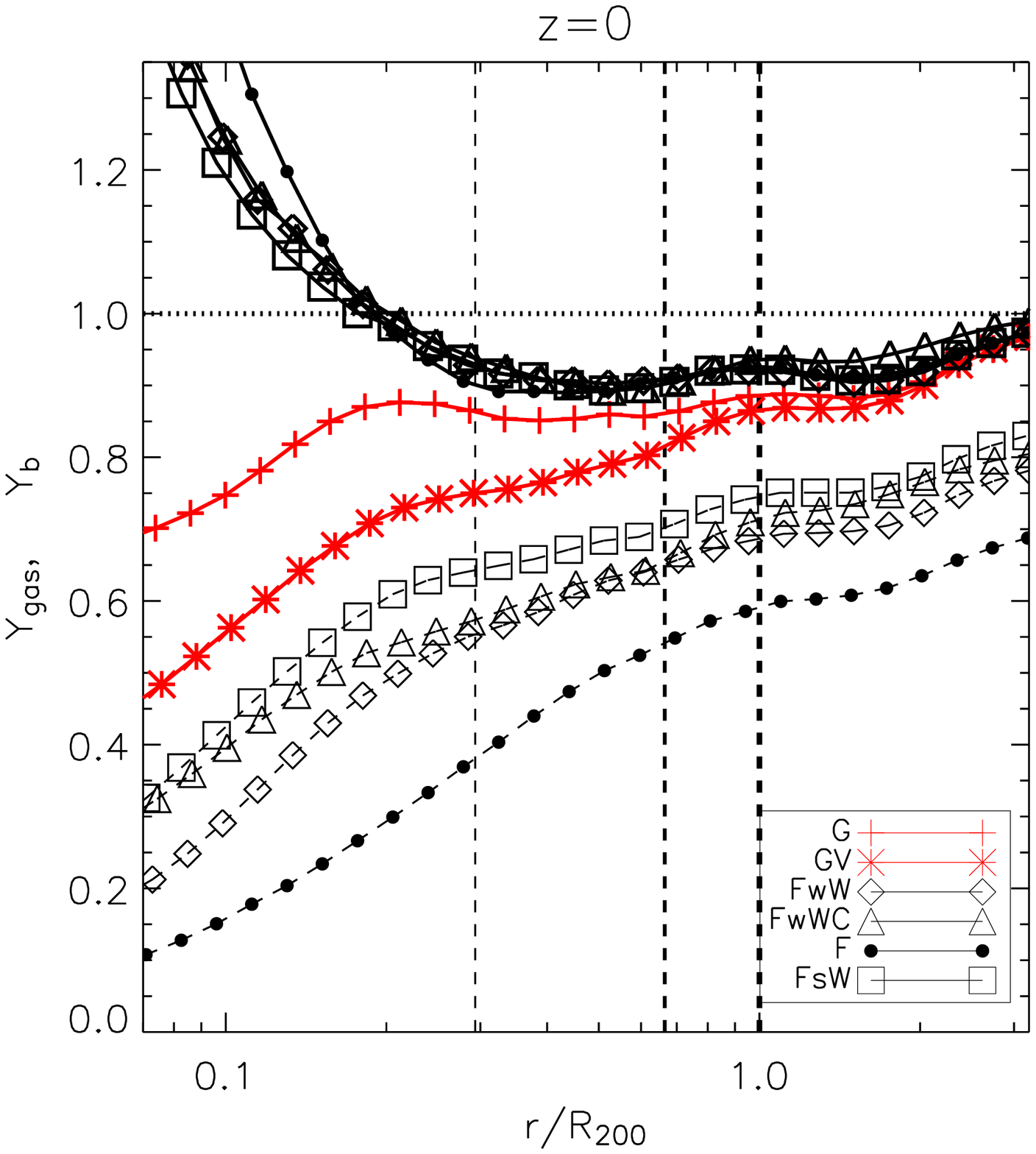,width=0.5\textwidth}
 \epsfig{figure=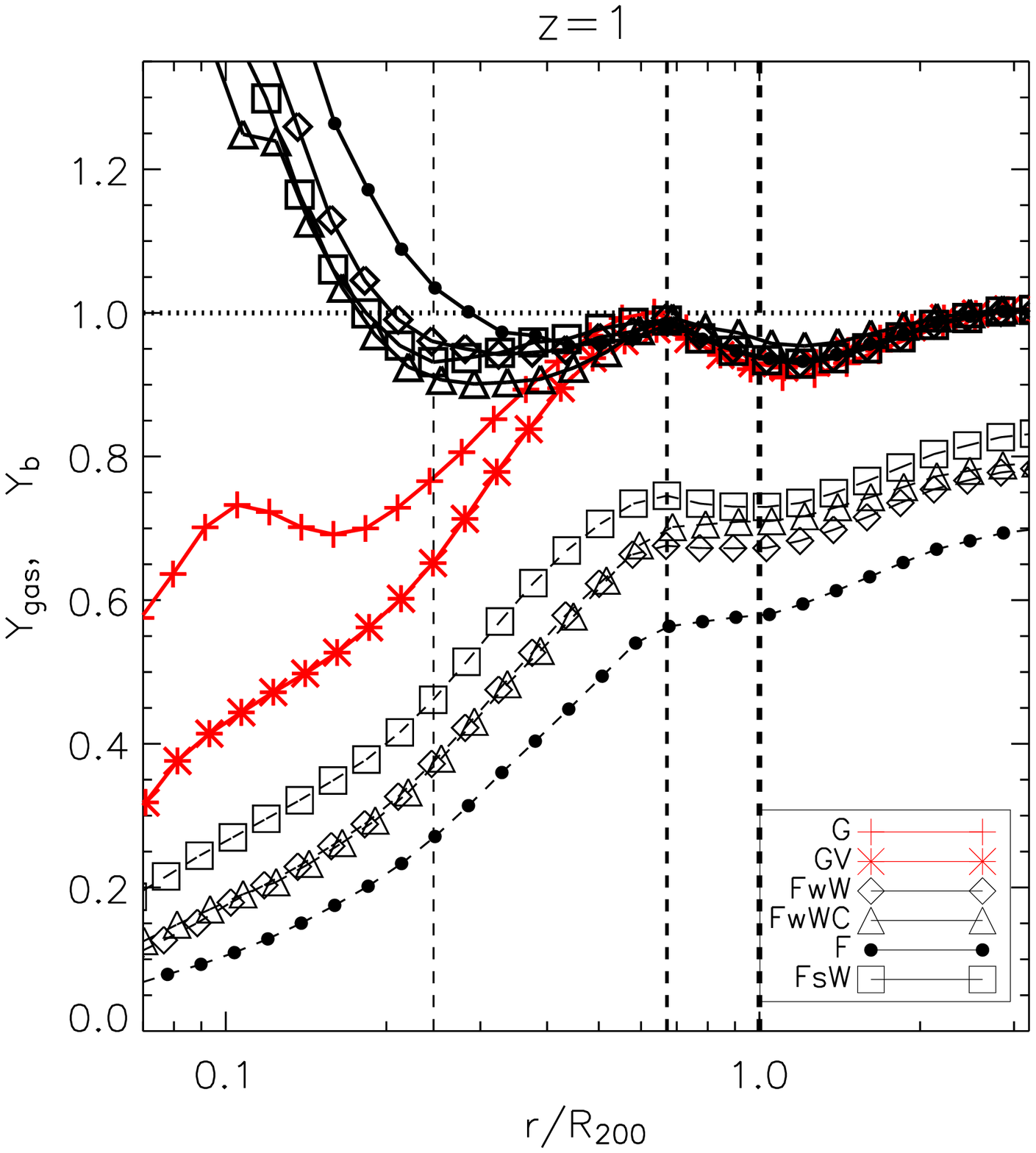,width=0.5\textwidth}
}
\caption{The radial distribution of the gas and baryon mass fractions
(in unit of the cosmic baryonic value) at $z=0$ and $1$
for one of the four massive re--simulated systems.
The vertical dashed lines indicate the location of $R_{2500}$, $R_{500}$
and $R_{200}$ for the {\it Gravitational heating only} case. 
}\label{fig:yb_r}
\end{figure}

In summary, this set of cluster was simulated 4 times, with different
kind of physics processes included:

\begin{itemize}
\item {\it Gravitational heating only (code=G)}.

\item {\it Gravitational heating only with low viscosity scheme (code=GV)}:
Like {\it G}, but using the alternative implementation of artificial
viscosity. In this simulation, galaxy clusters are found to have up to
30 per cent of their thermal energy in the turbulent motion of the ICM,
leading to a sizeable contribution of non--thermal pressure support in
the center of galaxy clusters;

\item {\it Cooling + Star Formation + Feedback with weak winds (code=FwW)}:
the wind speed is fixed at $\approx$340 km s$^{-1}$.

\item {\it Cooling + Star Formation + Feedback with weak winds and Conduction (code=FwWC)}:
Conduction efficiency set to be $1/3$ of the Spitzer rate.
\end{itemize}

In order to have under control the effect of changing the feedback
efficiency, 
one high--mass cluster was
further simulated with the following setups:

\begin{itemize}

\item {\it Cooling + Star Formation + Feedback with no winds (code=F)}:
Like {\it FwW}, but with winds switched off.

\item {\it Cooling + Star Formation + Feedback with strong winds (code=FsW)}:
Like {\it FwW}, but with wind speed increased to
$\approx$480 km s$^{-1}$, corresponding to a SN efficiency of unity.

%
 
\end{itemize} 

The center of each cluster is defined as the position of the
  particle having the minimum value of the gravitational
  potential. Starting from this position, we run a spherical
  overdensity algorithm to find the radius $R_{\Delta_c}$
  encompassing a given overdensity $\Delta_c$, with respect to the
  critical one at the redshift under exam, 
and the mass $M_{\Delta_c}$ enclosed within this radius.
In the present work, we consider values of the overdensity
$\Delta_c$ equal to $2500$, $500$ and $200$.
The corresponding radii relate to the virial radius, which defines a
sphere with virial overdensity (of $\approx 101$ at $z=0$ and
$\approx 157$ at $z=1$ for our cosmological model and with 
respect to the critical value), as  
$(R_{2500}, R_{500}, R_{200}) \approx (0.2, 0.5, 0.7) \times R_{\rm vir}$. 
For each cluster, the hot gas mass fraction and the stellar mass
fraction within a given radius $r$ are then calculated as $f_{\rm
gas}(<r) = M_{\rm gas}(<r) / M_{\rm tot}(<r)$ and $f_{\rm star}(<r) =
M_{\rm star}(<r) / M_{\rm tot}(<r)$, respectively.

\section{Results for the gas and stellar mass fractions}

For the sake of clarity, we define the quantities $Y_{\rm gas}$,
$Y_{\rm star}$ and $Y_{\rm b}$ as the ratios between
$f_{\rm gas}$, $f_{\rm star}$ and $f_{\rm b} = f_{\rm gas}+
f_{\rm star}$, and the cosmic value adopted in the present simulations,
$\Omega_{\rm b}/\Omega_{\rm m} = 0.13$.
Their mean values (and standard deviations of the observed
distributions, where available) at redshift
$z=0$, $0.3$, $0.7$ and $1$ are quoted in Table~\ref{tab:fgas}
for the different physical conditions considered.

We compare in Fig.~\ref{fig:ybar_m} the gas and baryon fraction as a
function of the virial mass, for both the simulated clusters extracted
from the cosmological box and for the subset of the re--simulated ones
with $M_{\rm 200} > 10^{15} h^{-1} M_{\odot}$. This plot demonstrates
that, when computed within the whole cluster virial region, there is
no relevant dependence upon mass of such fractions.  We remind the reader
that the two sets of clusters have been simulated for the same
cosmological model, the only difference being in the value of
$\sigma_8$, assumed to be 0.8 and 0.9 for the first and for the second
set, respectively.  In Fig.~\ref{fig:ybar_m}, the re-simulated
clusters indicated with a diamond correspond to the same simulation
physics as for the clusters extracted from the cosmological box. They
show a mean $Y_{\rm b}$ consistent with that measured for the
massive systems of the cosmological box, but with a slightly lower
value of $Y_{\rm gas}$. This is due to the increase in the cooling
efficiency with the amplitude of the power--spectrum, which produces
more evolved clusters (Borgani et al. 2005, in preparation).

\begin{figure*}
\hbox{
  \epsfig{figure=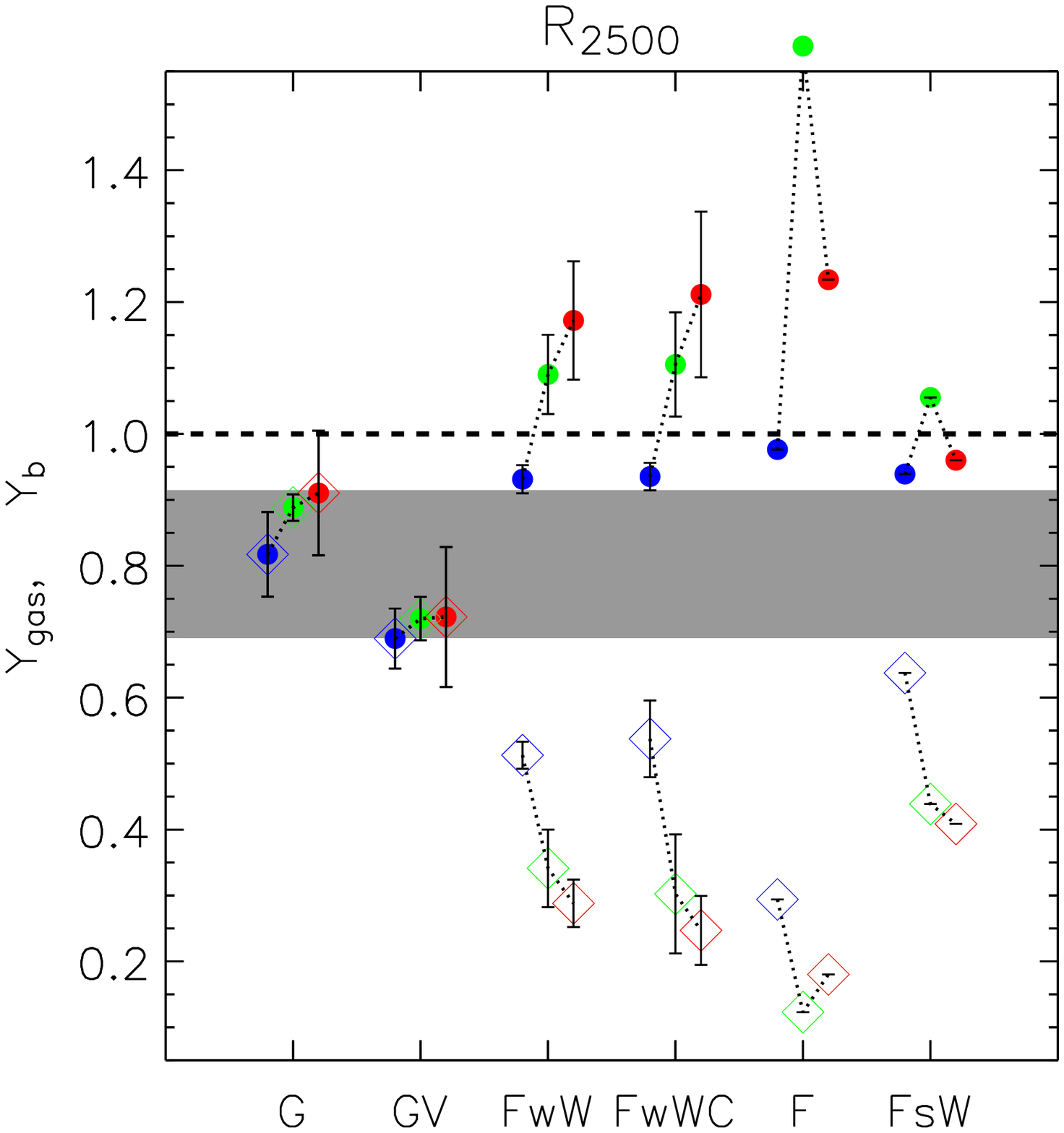,width=0.33\textwidth}
  \epsfig{figure=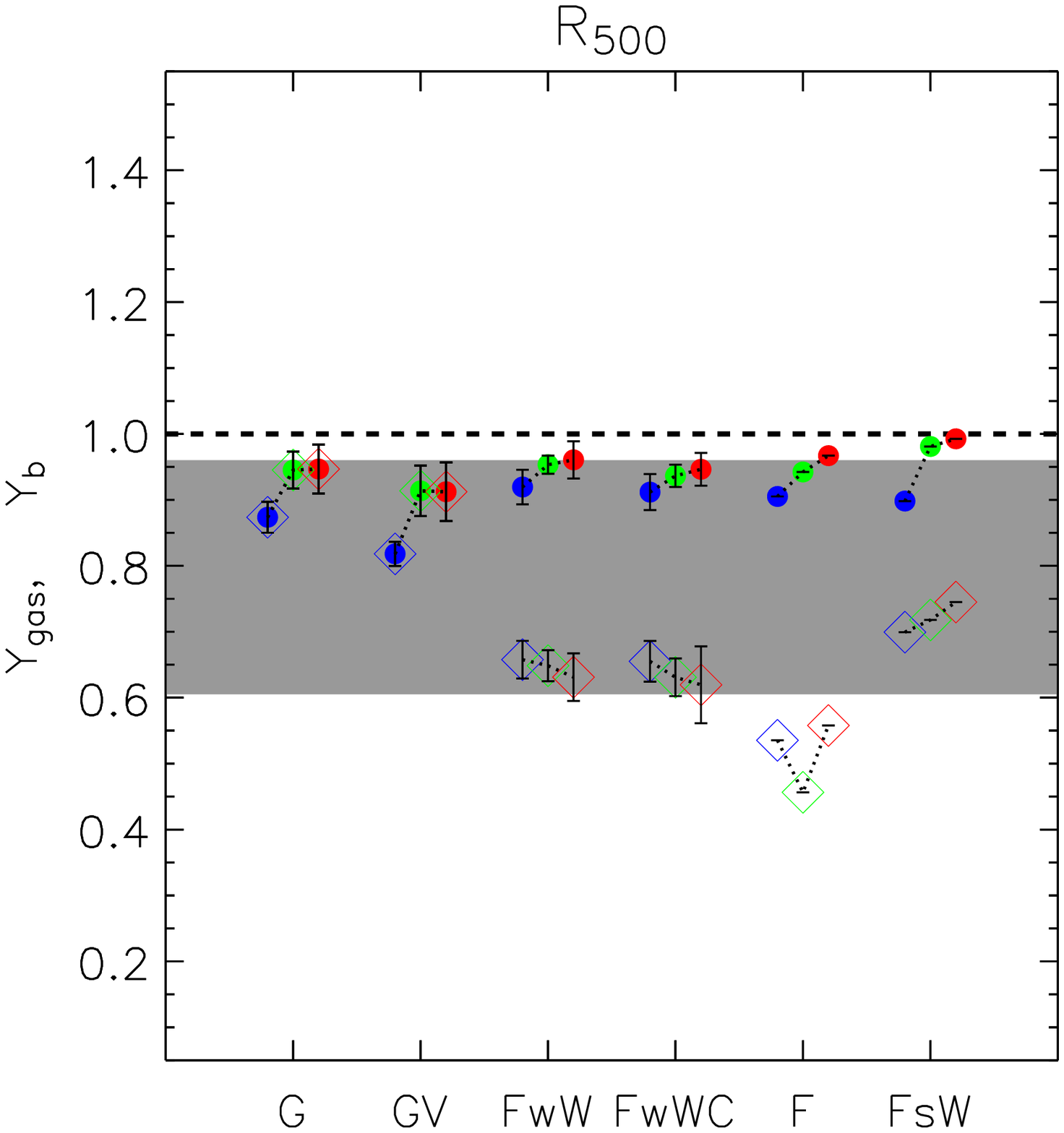,width=0.33\textwidth}
  \epsfig{figure=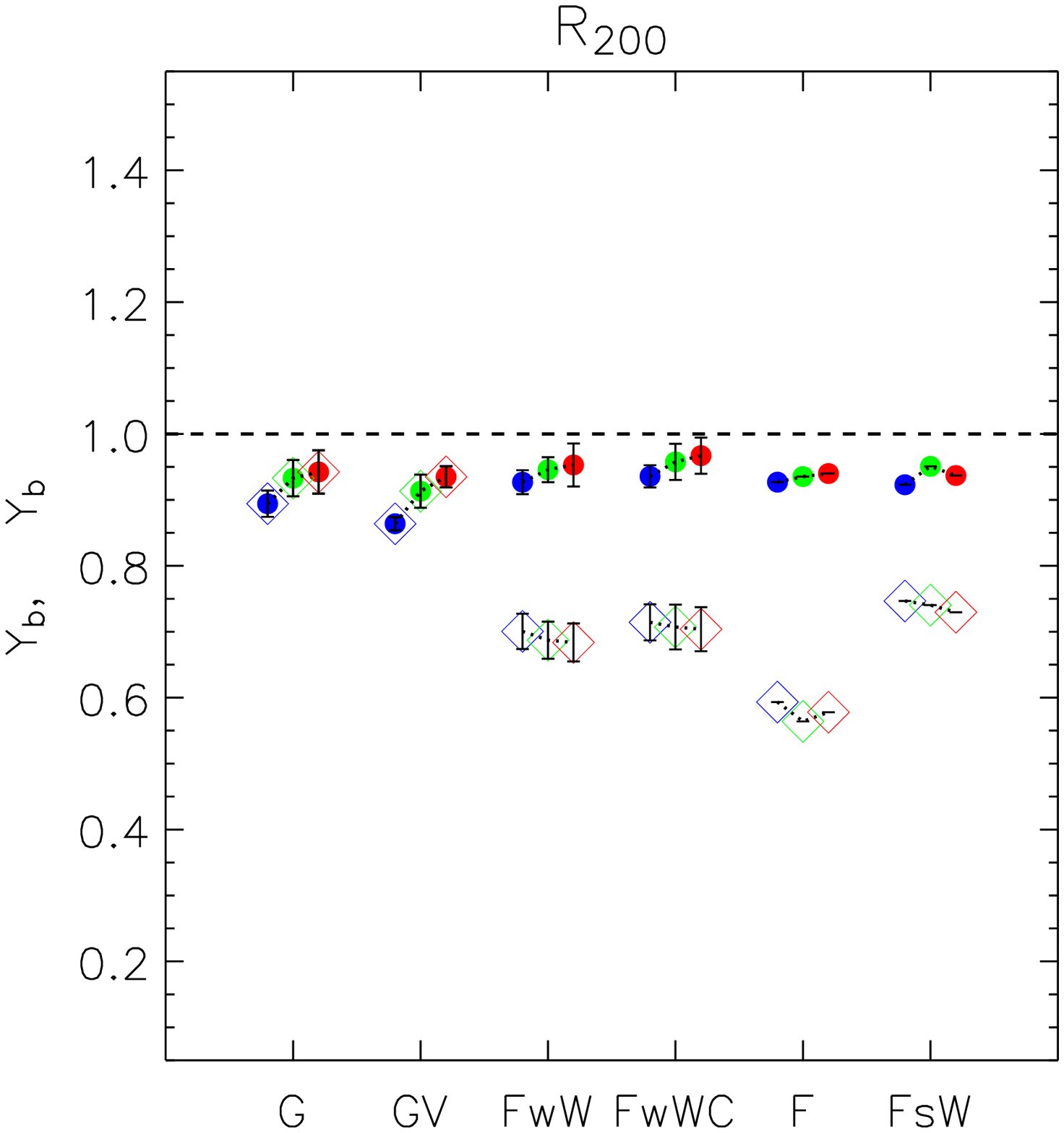,width=0.33\textwidth}
}
\caption{The gas ({\it diamonds}) and total baryon ({\it dots})
mass fractions, in unit of the cosmic baryonic
value at $R_{2500}$ (left panel), $R_{500}$ (central panel) and
$R_{200}$ (right panel).
For each physical case considered, we plot the mean and standard
deviation values measured at $z=0$, $0.7$ and $1$.  The shaded region
show the error-weighted mean and standard deviation of (1) $f_{\rm
gas}(R_{2500})$ estimated from 26 X-ray luminous galaxy clusters in
Allen et al. (2004, quoted in Table~2 for a LCDM universe), (2)
$f_{\rm gas}(R_{500})$ from 35 highly luminous ($L_{\rm X} \ga
10^{45}$ erg s$^{-1}$) objects in Ettori \& Fabian (1999). The
observational data are normalized to $\Omega_{\rm b} h^2 = 0.0214 \pm
0.0020$ (Kirkman et al. 2003), $H_0 = 70$ km s$^{-1}$ Mpc$^{-1}$ and
$\Omega_{\rm m} = 0.3$.  }\label{fig:ybar}
\end{figure*}

The radial distribution of the gas and baryon mass fractions
in these massive systems present quite a similar
behavior. For one representative object, we plot in
Fig.~\ref{fig:yb_r} these ratios as a function of radius, out to
$3 \times R_{200}$,
at $z=0$ and $1$, for the six physical schemes
adopted in our analysis.
The total baryon fraction measured in the radiative models 
is higher than the value obtained for the non--radiative runs 
(code$=G, GV$) within the virial radius and 
reaches the cosmic value at about $3 \times R_{200}$.
The gas fraction is larger when the winds are
stronger as a consequence of their effect in preventing gas removal
from the hot phase.
At $z=0$, it increases typically with radius reaching 
50 and 80 per cent of the value measured at $R_{200}$
at $r \la 0.1 R_{200}$ and at $\sim 0.3 R_{200}$,
respectively.
At $z=1$, the gas fraction is less concentrated: the radii
at which the 50 and 80 per cent of $f_{\rm gas}(<R_{200})$
is reached are a factor 1.4 and 2 larger than at $z=0$.

The quantities $Y_{\rm gas}$ and $Y_{\rm b}$ at $R_{2500}$, $R_{500}$
and $R_{200}$, as a function of redshift and physics included in the
simulations, are plotted in Fig.~\ref{fig:ybar}.  
In the inner cluster regions, the dissipative action of radiative
  cooling enhances the average $Y_{\rm b}$ to super-cosmic values at high
  redshift.  At late times cooling is less efficient, and $Y_{\rm b}$ declines,
  although to values ($\sim 0.9$ at $z=0$) that remain higher than those
  of the non-radiative runs.
A smaller scatter
and more widespread agreement among the different physical regimes are
instead found in the outskirts ($r \ga R_{500}$).  The gas fraction
within $R_{2500}$ is about $0.3$ times the cosmic value at $z=1$ and
$0.6$ at $z=0$, whereas is more tightly distributed around $0.6-0.7$
at larger radii, with evidence of larger values in the presence of
strong winds.  We also compare in Fig.~\ref{fig:ybar} our simulation
results with the observed $f_{\rm gas}$ distribution in highly X-ray
luminous clusters at $R_{2500}$ (from Allen et al. 2004) and at
$R_{500}$ (from Ettori \& Fabian 1999). 
Simulations clearly indicate a sizeable underestimate of the hot 
baryons budget, both at $R_{2500}$ and at $R_{500}$. 
When extra physics is added to the action of
gravitational heating, lower hot gas fractions result.
The discrepancy with the inferred
observed fraction signals the existence of systematic errors,
either in our physical treatment, or in estimates of the observed
fraction, or possibly both. 
It is worth noticing that total mass estimates, for instance, suffer
from systematic differences when measured from X-ray analysis and 
from dark matter particles in simulations, mainly owing to
bias in the X-ray spectral temperature measurements 
(see, e.g., Mazzotta et al. 2004, Vikhlinin 2005 and, 
specifically related to the systematics in X-ray mass estimates, 
Rasia et al. 2005).  
The observed gas fraction measurements are well represented by
the values measured in the non--radiative simulations also in the inner
regions, where variations by $\la 15$ per cent due to the action
of non-thermal pressure support (induced by the reduced viscosity
scheme) encompass the observed variance.

\begin{figure*}
\hbox{
  \epsfig{figure=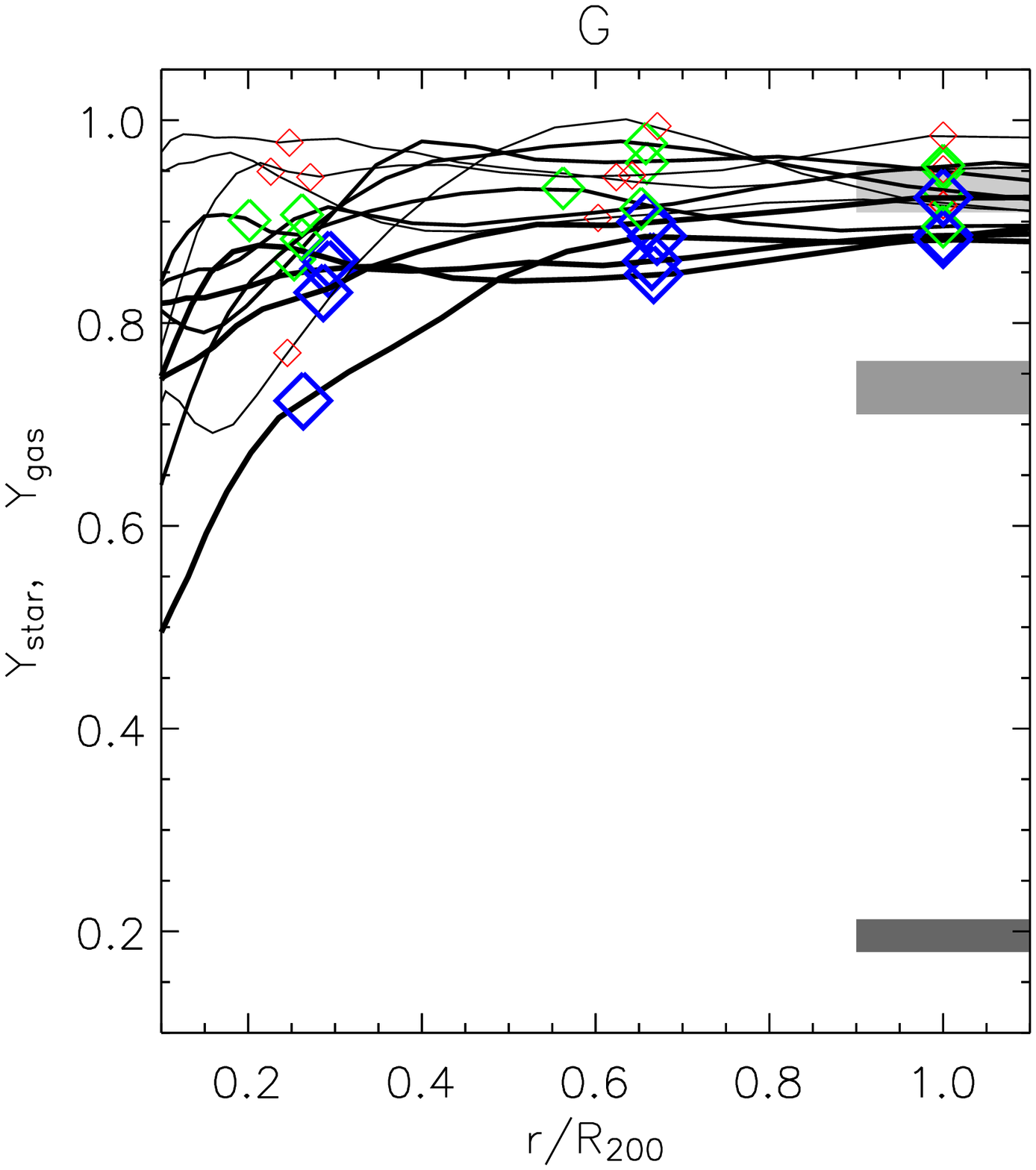,width=0.43\textwidth}
  \epsfig{figure=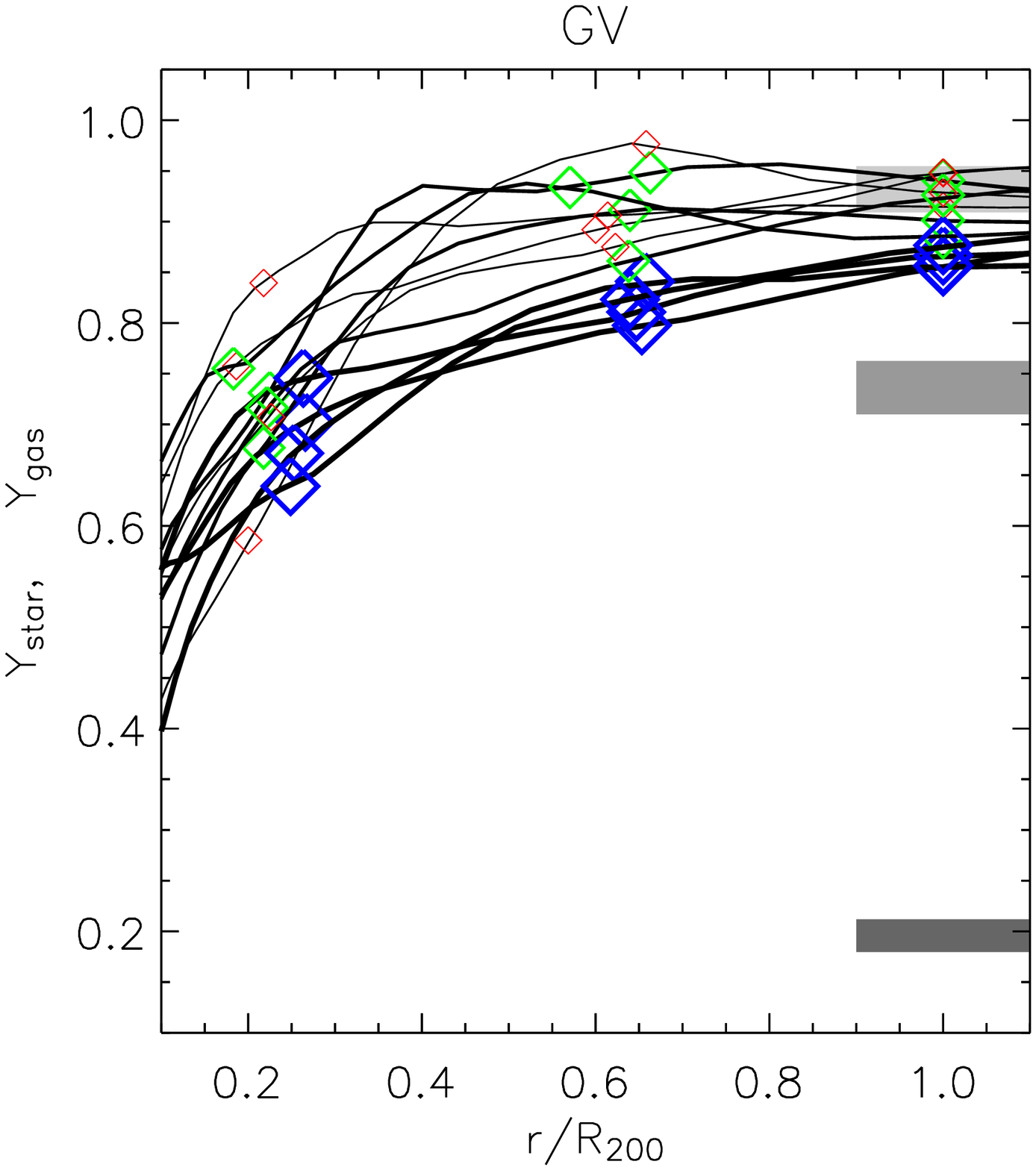,width=0.43\textwidth}
} \hbox{
  \epsfig{figure=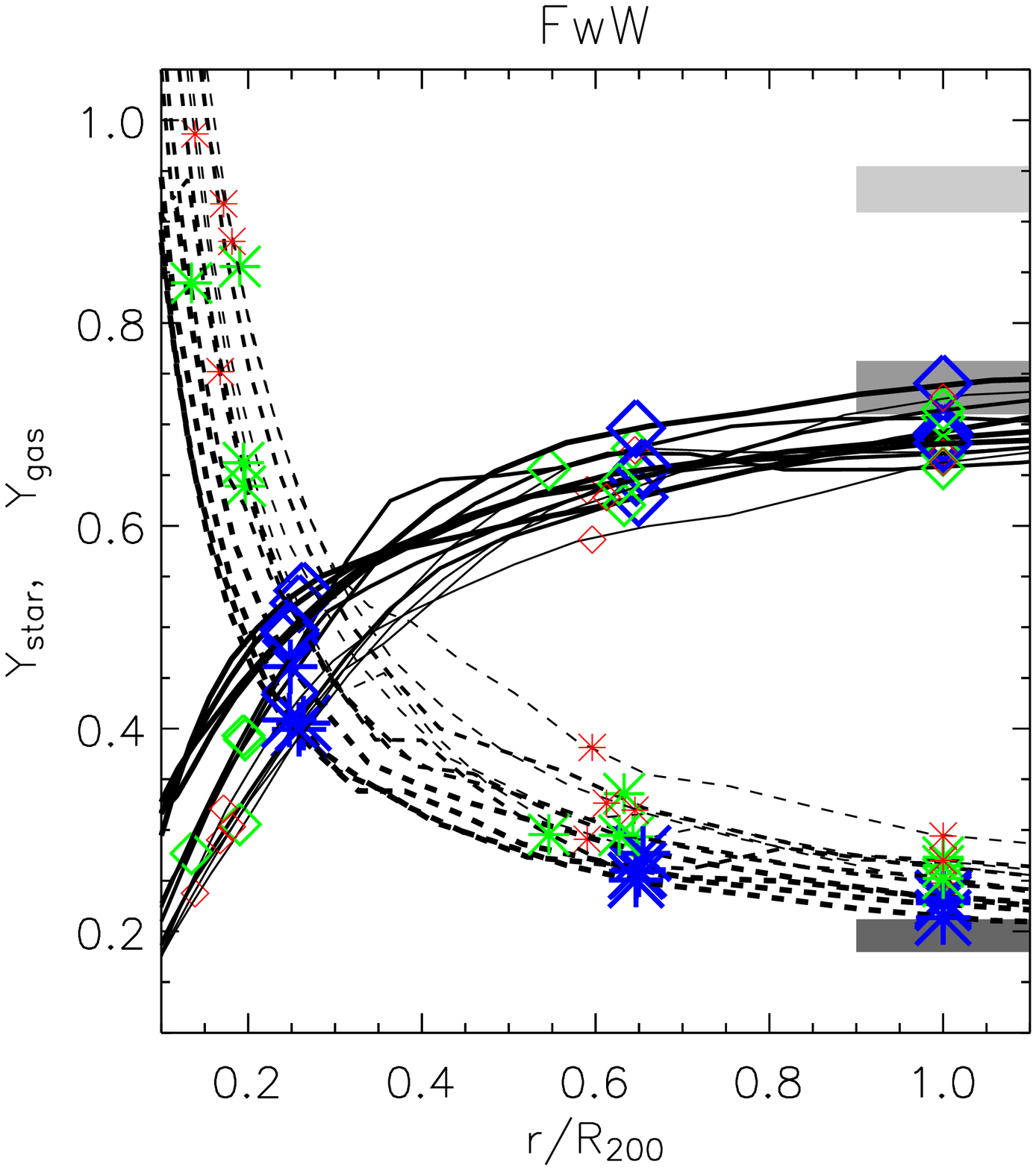,width=0.43\textwidth}
  \epsfig{figure=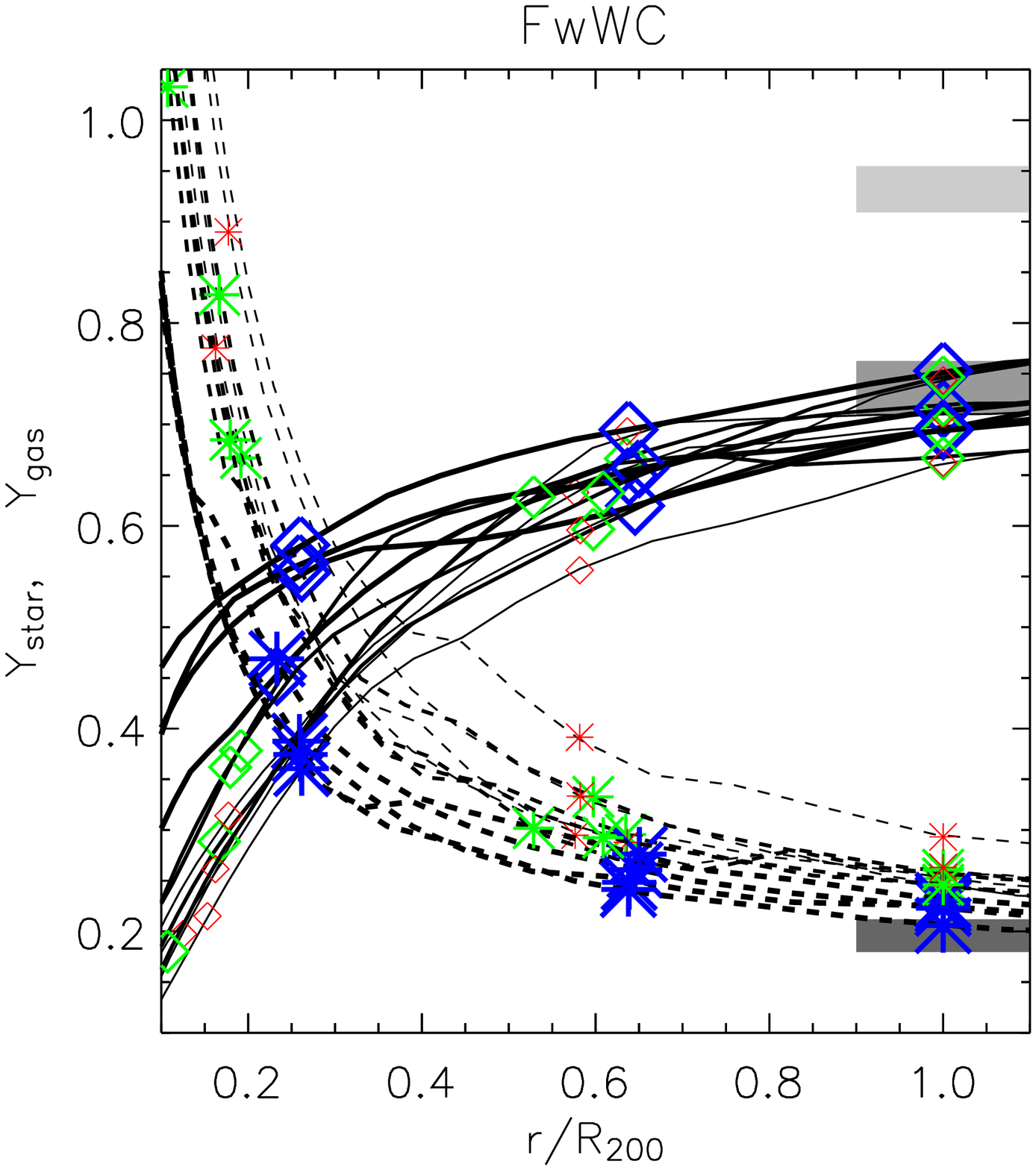,width=0.43\textwidth}
}  \hbox{
  \epsfig{figure=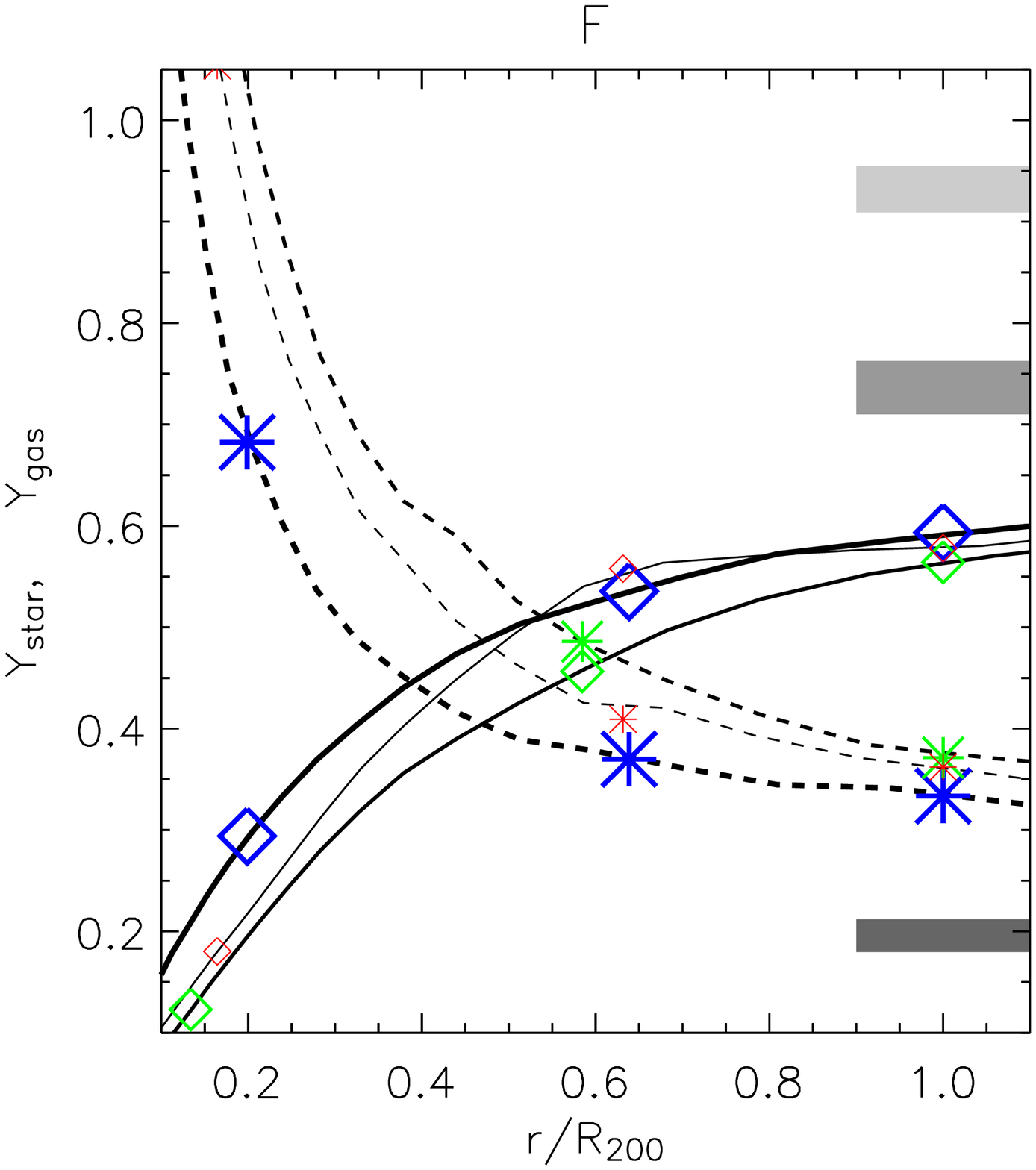,width=0.43\textwidth}
  \epsfig{figure=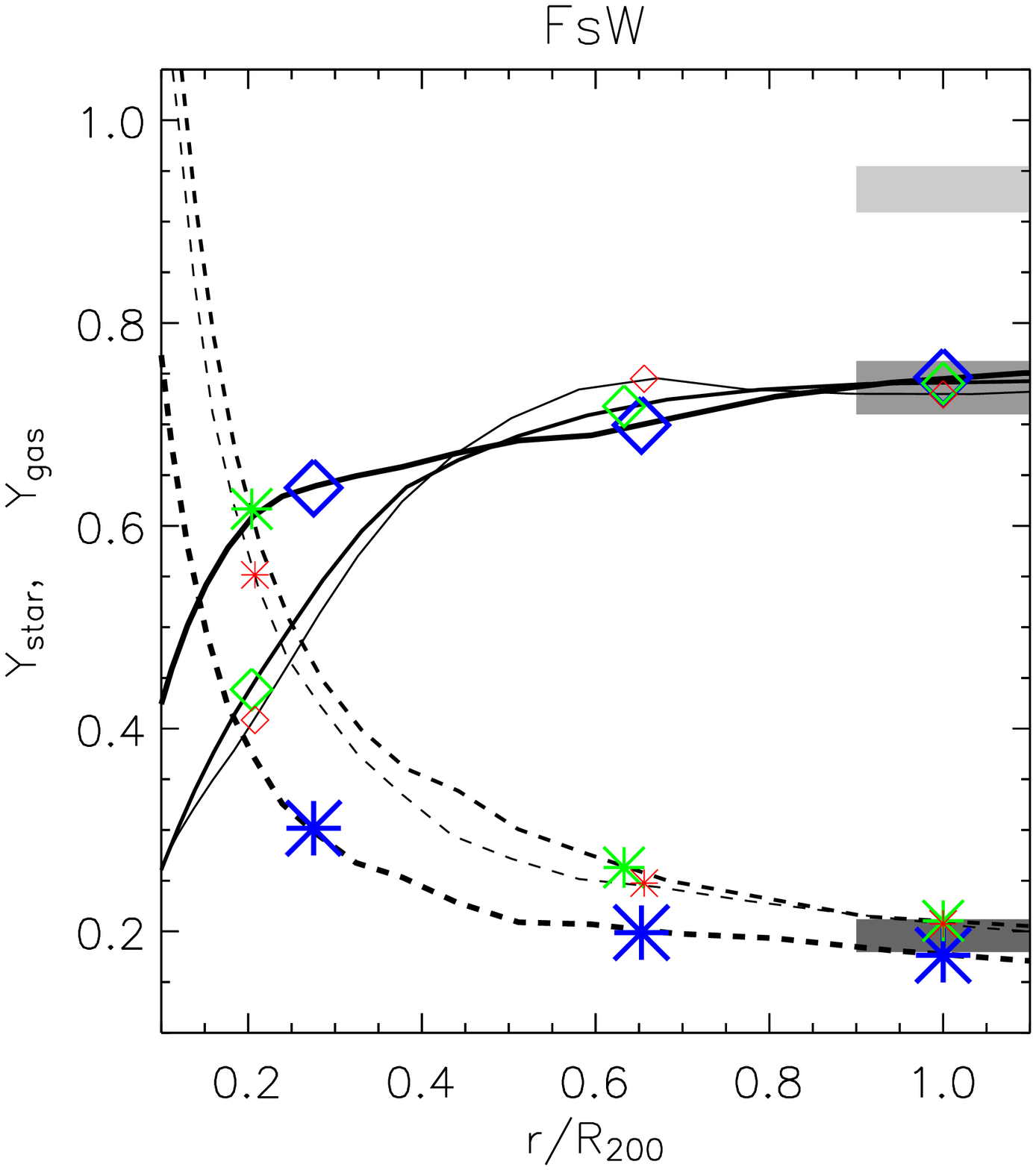,width=0.43\textwidth}
}
\caption{The comparison between the outputs of the 6 different
physical schemes investigated. The gas ({\it solid line}) and stellar
({\it dashed line}) mass fractions, normalized to the cosmic value,
are plotted at $R_{2500} (\approx 0.3 R_{200})$, $R_{500} (\approx 0.7
R_{200})$ and $R_{200}$.  Their evolution with redshift is indicated
by the thickness of the lines (from thickest line/larger symbols to
thinnest line/smaller symbols: $z=0$, $0.7$, $1$).  The shaded regions
indicate the $1 \sigma$ range of $Y_{\rm star}$, $Y_{\rm gas}$ and
$Y_{\rm b}$ measured at $R_{200}$ and $z=0$ for the cluster set
extracted from the cosmological box.  } \label{fig:fgas}
\end{figure*}

\begin{figure*}
\hbox{
  \epsfig{figure=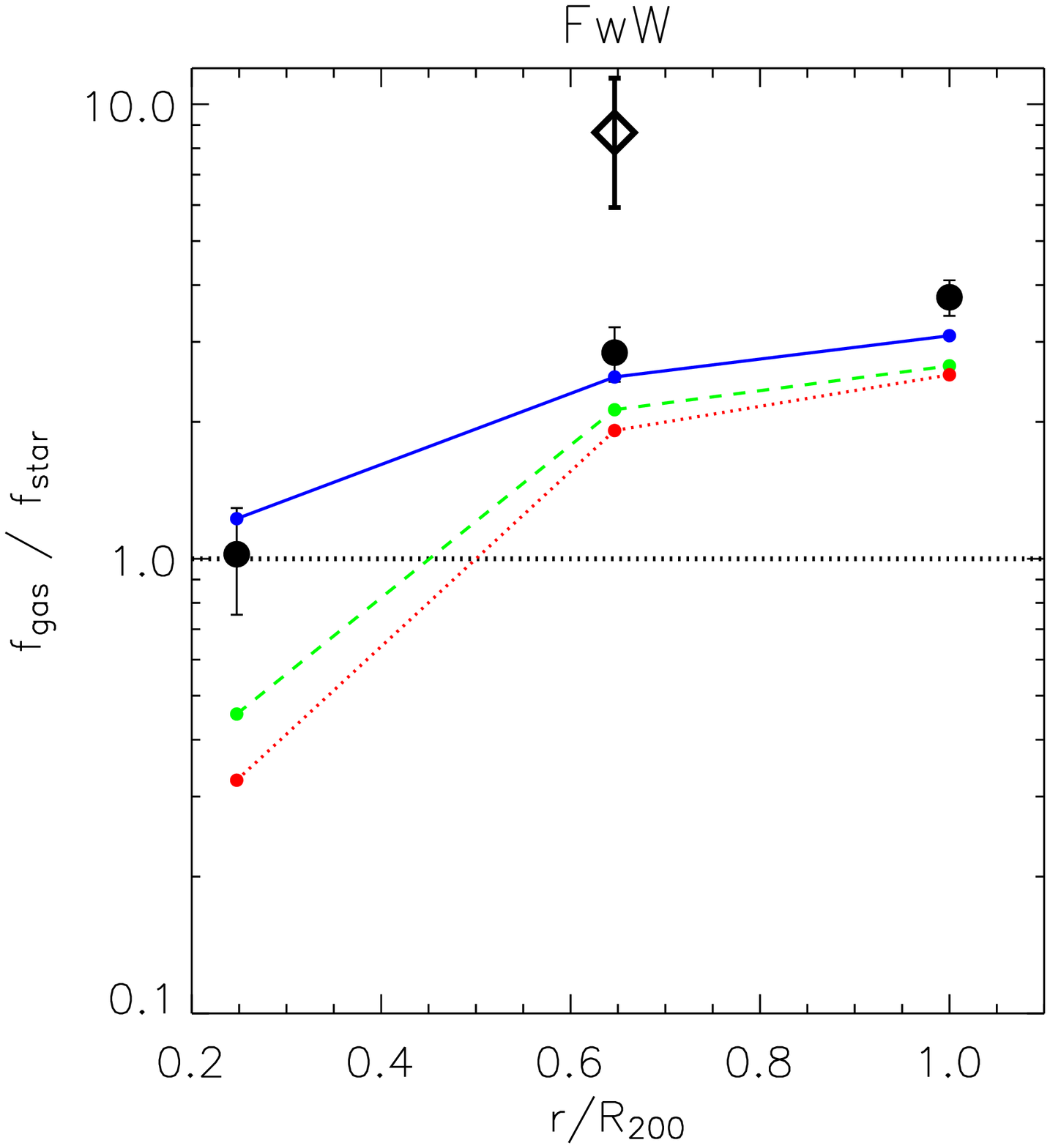,width=0.45\textwidth}
  \epsfig{figure=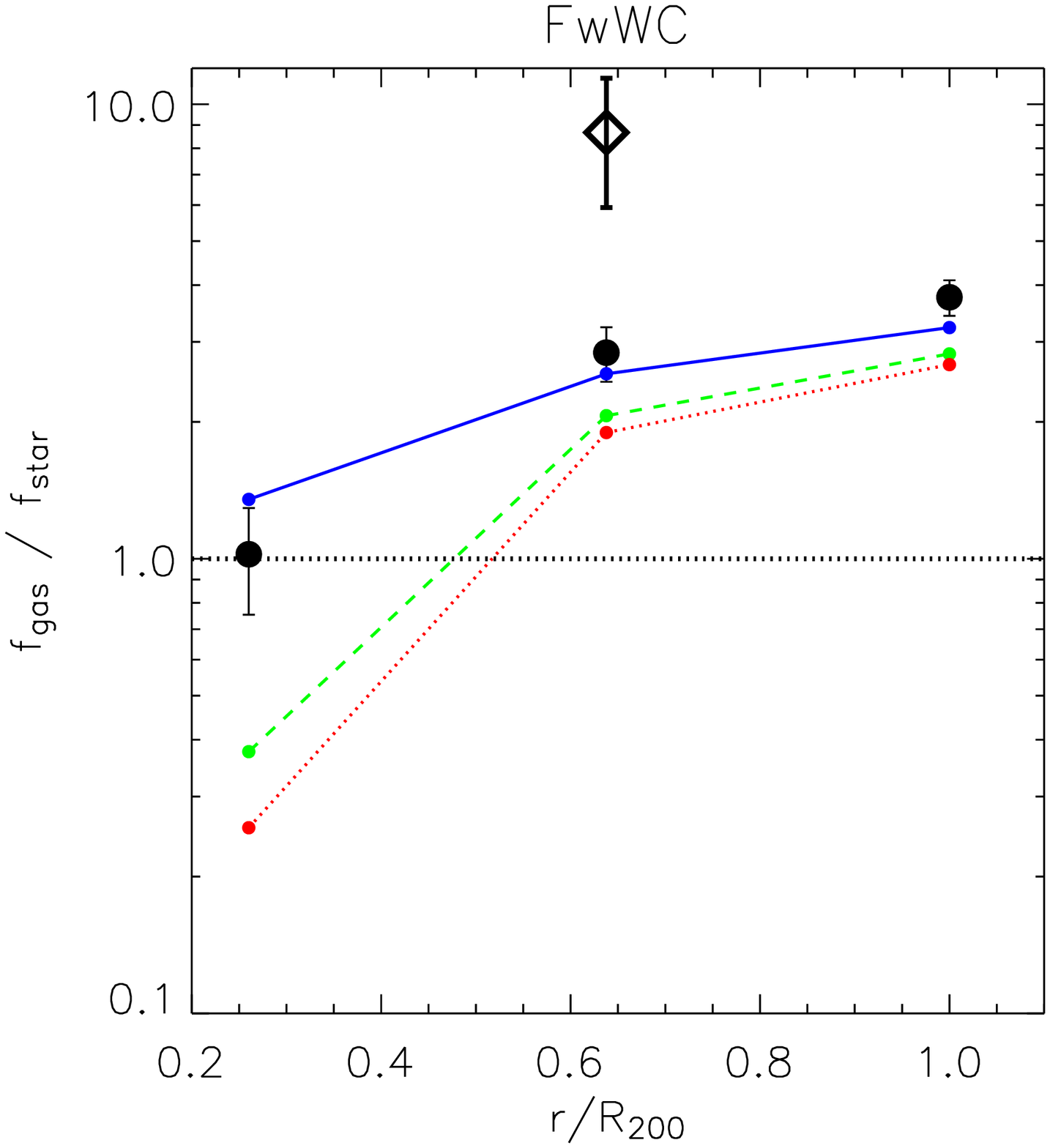,width=0.45\textwidth}
}  \hbox{
  \epsfig{figure=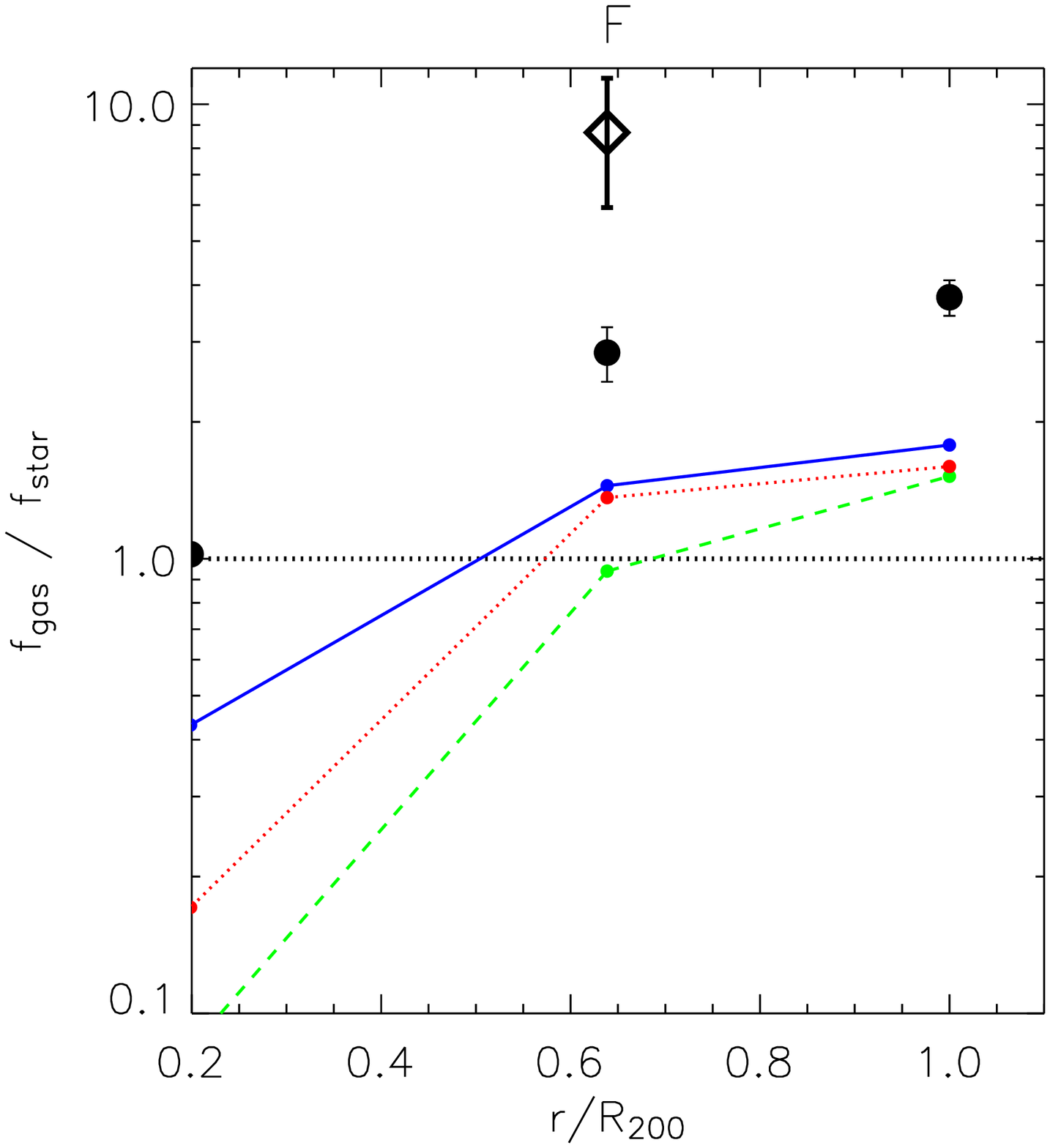,width=0.45\textwidth}
  \epsfig{figure=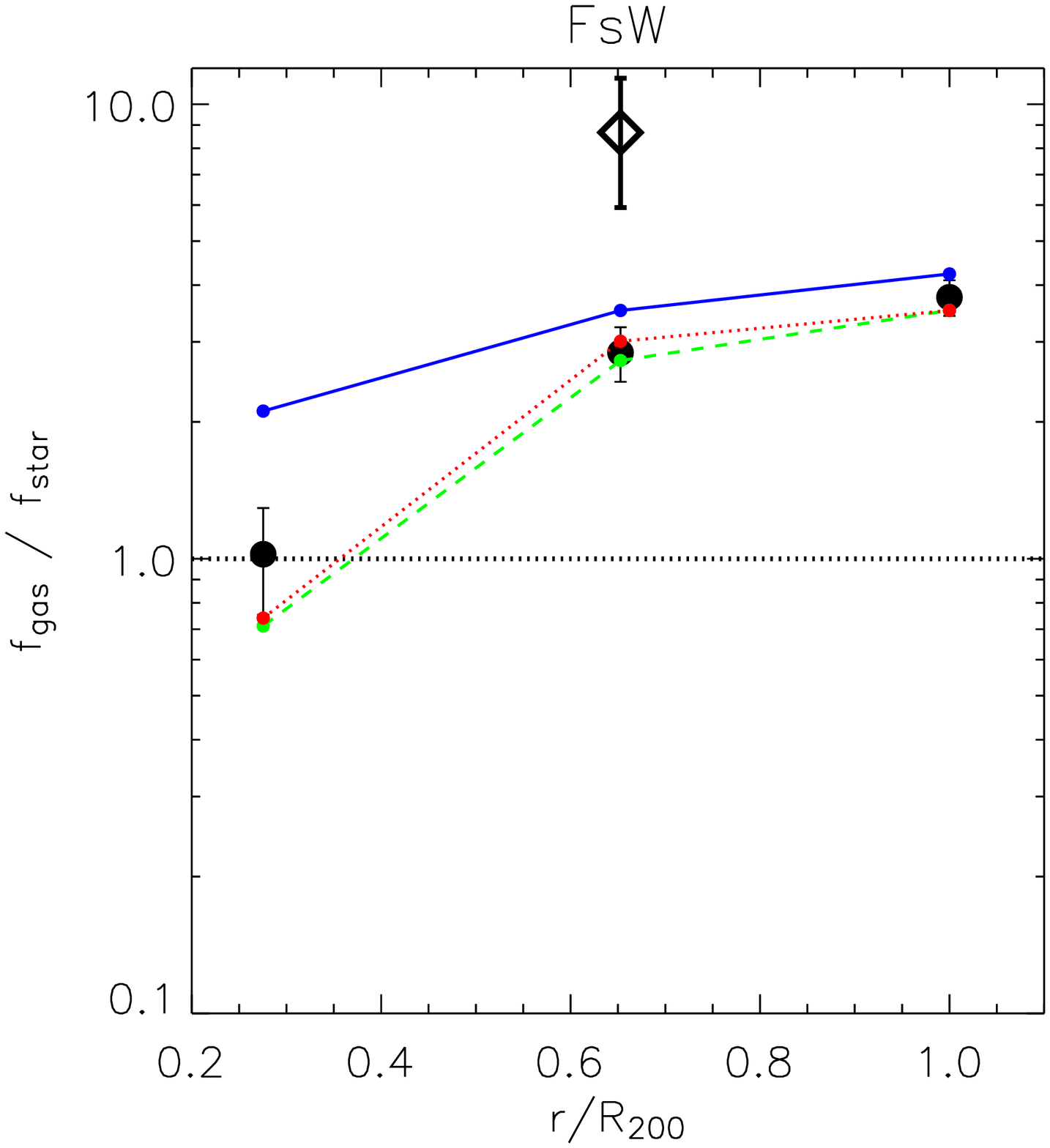,width=0.45\textwidth}
}
\caption{Ratios between 
the cumulative gas and stellar mass 
fractions within a given radius as a
function redshift
({\it solid line}: $z=0$; {\it dashed
line}: $z=0.7$; {\it dotted line}: $z=1$) for the different physics
included in the simulations.  The large dots show the mean value (and
relative standard deviation) obtained from the cosmological sample.
The diamonds indicate the ratios expected, for a given $M_{500}$, from
the equation~10 of Lin, Mohr \& Stanford (2003), based on
near--infrared observations of massive X-ray galaxy clusters.  }
\label{fig:fgas_r}
\end{figure*}

The values of, and the relation between, $Y_{\rm gas}$ and $Y_{\rm
star}$ are shown in Fig.~\ref{fig:fgas} and
\ref{fig:fgas_r}.  At $R_{2500}$, the estimate of $Y_{\rm b}$ is
strongly affected by the physics involved in the accumulation of the
cluster baryons, with super-cosmic values mainly due to a very high
star-formation efficiency ($Y_{\rm star} > 0.5$), particularly at high
redshift (see Fig.~\ref{fig:fgas}). The effect of this cooling excess
is to sink more gas in the core to maintain the pressure support, thus
making the overall baryon fraction super-cosmic and the gas fraction
less concentrated at high redshift. At $R_{500}$ and $R_{200}$, the
dispersion in the estimate of $Y_{\rm b}$ is 2-3 per cent of the
measured mean (Fig.~\ref{fig:ybar}).  We explain the reduction in the
scatter at large radii by the common history that the baryons
experience in our simulations when the properties are averaged over a
volume large enough not to be dominated from the physics of the
core. With respect to the mean values measured at $R_{2500}$, $Y_{\rm
star}$ decreases by a factor $\ga 2$ and $Y_{\rm gas}$ increases by
more than 30 per cent, with peaks of 2-3 when no winds, or weak winds
combined with the effect of conduction, affect the ICM physics
(Fig.~\ref{fig:fgas_r}). The ratio $f_{\rm gas}/f_{\rm star}$, that
is $\la 1$ at $R_{2500}$ and $z\ga 0.3$, increases with both radius
and redshift and becomes $\simeq 3$ at $R_{200}$ and $z=0$ in the
presence of winds (see Fig.~\ref{fig:fgas_r}). Once again, the absence
of galactic winds increases the efficiency of star formation and
leaves the $f_{\rm gas}/f_{\rm star}$ ratio below $2$ within the
virial radius. These values are well above the observed constraints,
also plotted in Fig.~\ref{fig:fgas_r}.  Lin, Mohr \& Stanford (2003)
compare the gas mass measured in nearby X-ray bright galaxy clusters
with the stellar masses evaluated from K-band luminosities of the
member galaxies.  From their estimates, converted to the total mass
measurements in Arnaud et al.  (2005), we infer an average $M_{\rm
gas}/ M_{\rm star} \approx 8.7$ (r.m.s. $=2.7$) at $R_{500}$ in
systems with gas temperatures larger than 3 keV, that is still a
factor between 2.5 and 6 larger than what obtained in our simulated
objects (Fig.~\ref{fig:fgas_r}), thus witnessing the presence of
significant overcooling.

\section{Comparison to previous analyses}

While the comparison between results from different non--radiative
simulations is relatively straightforward, when extra-physics is
included, such as radiative cooling, star formation and stellar
feedback, the results on the hot and cool baryons distributions is
sensitive to the implemented scheme. Therefore, this difference has to
be taken into account in performing such a comparison.

In their non--radiative simulations, Eke, Navarro \& Frenk (1998)
found that the gas fraction within $R_{\rm vir}$ is, on average, 87
per cent of the cosmic value, that is reached at about $3 R_{\rm
vir}$.  For dynamically relaxed systems, they measure $Y_{\rm gas} =
0.824 \pm 0.033$ at $r = 0.25 R_{\rm vir} \approx R_{2500}$ (as quoted
in Allen et al. 2004).  No evident evolution is present between $z=1$
and $0$.  A comparable value was obtained as average of the baryon
fraction measured in a set of non--radiative simulations of one single
cluster presented in the Santa Barbara Comparison Project (Frenk et
al. 1999), with $f_{\rm b} / (\Omega_{\rm b}/\Omega_{\rm m})$ of
$0.92$ (rms: $0.07$) at the virial radius.  Similar results have been
obtained also in other SPH simulations (e.g. Bialek et al. 2001,
Muanwong et al. 2002).

These results agree with our estimates for 
our simulations of massive galaxy clusters with gravitational heating
only: $Y_{\rm b}(<R_{200}) = 0.89 \pm 0.02$ ($0.90 \pm 0.03$ at
$R_{\rm vir}$).  A slight increase at high redshift is however measured
($Y_{\rm b} = 0.94 \pm 0.03$ at $z=1$).  On the contrary, the baryon
fraction decreases moving inwards, with $Y_{\rm b}(<R_{500}) = 0.87$
and $Y_{\rm b}(<R_{2500}) = 0.82$

Kravtsov, Nagai \& Vikhlinin (2005) studied the baryon fraction in nine
galaxy clusters spanning a decade in mass and simulated with the
Eulerian adaptive mesh refinement N-body+gasdynamics {\small ART} code, 
for both non--radiative and radiative cases.
For non--radiative simulations, they measure $Y_{\rm b} \approx
1$ (at $r \ga 3 R_{\rm vir}$), $0.97 \pm 0.03$ (at $R_{\rm vir}$),
$0.94 \pm 0.03$ (at $R_{500}$) and $0.85 \pm 0.08$ (at $R_{2500}$),
that are consistently higher than our values by few per cent.
This systematic difference by $\approx 5$\% at $r \ga R_{2500}$
has been already shown by the comparison between the {\small ART} 
and {\tt GADGET} codes discussed in Kravtsov et al. (2005) 
and, even though smaller than the $10$\% differences measured
in the Santa Barbara cluster comparison project (Frenk et al. 1999)
between the gas fraction obtained from Eulerian and SPH codes,
is still relevant when using simulations to 
calibrate systematics in the estimate of the baryon fraction 
in clusters.


When the radiative cooling and other physical processes are turned on,
the number of reliable comparisons that can be done is reduced.  In
the radiative simulations of Muanwong et al. (2002), $Y_{\rm gas}$ is
between $0.6-0.7$ in high$-M$ systems and $Y_{\rm b} \approx 0.9$ at
the virial radius.  When pre--heating with an extra energy of 1.5 keV
per particle at $z=4$, these authors find $Y_{\rm gas} \sim Y_{\rm b}
\approx 0.8-0.9$.  Kravtsov et al. (2005) measure the gas fraction to
rise from $0.46 \pm 0.06$ (at $R_{2500}$) to $0.65 \pm 0.06$ (at
$R_{\rm vir}$), whereas $Y_{\rm b}$ remains always around $1$ (from
$1.22 \pm 0.11$ at $R_{2500}$ to $1.02 \pm 0.02$ at $R_{\rm
vir}$). These values are roughly in agreement with our results, in
particular for the set of simulated objects with weak winds: at $z=0$,
$Y_{\rm gas}=0.51 \pm 0.02$ at $R_{2500}$ and $0.70 \pm 0.03$ at
$R_{200}$, while $Y_{\rm b}$ is flat around $0.93 \pm 0.02$ and never
super-cosmic, i.e.  $Y_{\rm b} < 1$.  Different numerical
implementations of the cooling, star--formation and feedback processes
are expected to contribute to this systematic difference between the
predictions of {\small ART} and {\tt GADGET} simulations.  These
differences are, in fact, more significant in the radiative runs and
depend on the amount of baryons cooled in stars (see also Kravtsov et
al. 2005). Such differences, which are numerical in origin, should
definitely be considered as theoretical uncertainties, when using
simulations to calibrate systematic biases in the observational
estimate of the baryon fraction within clusters. The action of
stronger winds is to increase the gas fraction, whereas the absence of
winds reduces $Y_{\rm gas}$ to $0.29$ and $0.59$ at $R_{2500}$ and
$R_{200}$, respectively.

\section{Implications for the constraints on cosmological parameters}

Our results have a direct implication on the systematics that affect
the constraints on the cosmological parameters obtained through the
cluster baryon mass fraction (e.g. White et al. 1993, Evrard 1997,
Allen et al. 2002, Ettori et al. 2003, Ettori 2003, Allen et
al. 2004).  We remind that, once a representative gas fraction, 
denoted here $\hat{f}_{\rm gas}$, is directly measured 
from X-ray observations and a statistical relation between the average 
$\hat{f}_{\rm star}$ and $\hat{f}_{\rm gas}$ is adopted, 
the cosmic mass density parameter can be then evaluated as
\begin{equation}
\Omega_{\rm m} = \frac{Y_{\rm b} \ \Omega_{\rm b}}{\hat{f}_{\rm gas}
\left( 1 + \hat{f}_{\rm star}/\hat{f}_{\rm gas} \right)},
\label{eq:cosmo} 
\end{equation}
where the ``hat" indicates the observed quantities and 
the cosmic baryon density $\Omega_{\rm b}$ is assumed from
primordial nucleosynthesis calculations or the measured anisotropies
in the cosmic microwave background.  In recent years this method has
been also extended to the measure of the dark energy density parameter,
($\Omega_{\Lambda}$, $w$; see, e.g., Allen et al. 2002, Ettori et
al. 2003, Allen et al. 2004) under the assumption that the gas
fraction remains constant in redshift
(Sasaki 1996, Pen 1997). Since the gas fraction
scales with the angular diameter distance as $f_{\rm gas} \propto d_{\rm
A}^{1.5}$, the best choice of cosmological parameters is defined as
the set of values that minimizes the $\chi^2$ distribution of the
measured gas fraction at different redshifts
with respect to the reference value:
\begin{equation}
f^{\Lambda CDM}_{\rm gas} = \frac{Y_{\rm b} \ \Omega_{\rm b}
\Omega_{\rm m}^{-1}} {\left( 1 + \hat{f}_{\rm star}/\hat{f}_{\rm gas}
\right)} \left[ \frac{d_{\rm A}(z; \Omega_{\rm m}=0.3;
\Omega_{\Lambda}=0.7)}{d_{\rm A}(z; \Omega_{\rm m}; \Omega_{\Lambda})}
\right]^{1.5}.
\label{eq:fgas} 
\end{equation}

Despite its conceptual simplicity and straightforward application,
this method makes some assumptions that have to be tested before
the error bars estimated for the matter and dark--energy density parameters
can be accepted as robust and reliable determination of both
statistical and systematic uncertainties.  In the present discussion,
we highlight two of the assumptions generally adopted, but never
verified: (1) the mean value of $Y_{\rm b}$ does not evolve with
redshift, (2) a fixed ratio between $f_{\rm star}$ and $f_{\rm gas}$
holds in a cluster at any radius and redshift.  As we have shown here,
both these assumptions are not valid in our simulated dataset whatever
is the physics included in the simulations, in particular when
considering the inner part of the clusters.  Allen et al. (2004) use
the simulation results by Eke et al. (1998) to fix $Y_{\rm b}=0.824
\pm 0.033$ at $r \approx R_{2500}$ for their sample of {\it Chandra}
exposures of the largest relaxed clusters with redshift between $0.07$
and $0.9$. We notice, for instance, that, while this value is in
agreement with our simulation results at $z=0$ in the runs with
{\it Gravitational heating only} ($Y_{\rm b}=0.82 \pm 0.06$), it is
definitely lower than what we estimate at higher redshift (e.g. $Y_{\rm
b}=0.86, 0.89, 0.91$ at $z=0.3, 0.7, 1$, respectively).
This increase of $Y_{\rm b}$ with redshift is the consequence of the
different accretion pattern of shock-heated baryons at different epochs. 
At later times, accreting gas had more time to be pre-shocked into filaments. 
As a consequence, they have a relatively higher entropy, thus relatively 
increasing the radius (in unit of the virial radius) where
accretion shocks take place. 
This is the reason why, as shown in Fig.~\ref{fig:yb_r}, at $z=1$
$f_{\rm b}$ reaches the cosmic value at relatevely smaller radii than at $z=0$.

Since the tighter cosmological constraints provided by the cluster gas
fraction alone are on $\Omega_{\rm m}$ (of the order of 16 per cent at
$1 \sigma$ level; e.g. Allen et al. 2004), we try to quantify the
effect of the variation of the baryonic components with the radius 
and the redshift on this estimate. 
To this purpose, we use equation~\ref{eq:cosmo}
and evaluate first how $\Omega_{\rm m}$ changes by varying $Y_{\rm b}$.
In runs with gravitational heating only, the increasing baryon fraction with 
redshift induces larger estimate of $\Omega_{\rm m}$ with respect to what
obtained from local measurements of $Y_{\rm b}$: 
\begin{equation}
\frac{ \Omega_{\rm m}^{\prime} - \Omega_{\rm m} }{\Omega_{\rm m}} =
 \frac{\Delta \Omega_{\rm m}}{\Omega_{\rm m}} = 
 \frac{Y_{\rm b}(<R_{\Delta}, z=z_o)}{Y_{\rm b}(<R_{\Delta}, z=0)} -1
\end{equation} 
is $+0.09$ at $R_{\Delta}=R_{2500}$ and $z_o=0.7$ for the case with 
gravitational heating only and $+0.11$ at $z_o=1$. 
(Here the prime symbol $^{\prime}$ indicates the corrected value with 
respect to the reference one).
Using instead the runs with reduced viscosity, the deviation decreases 
to about $+0.05$. As for the radiative runs, the bias is of the order 
of 20 per cent, that reduces to 2 per cent in presence of strong winds 
at $z=1$.  When outer cluster regions are mapped (i.e. $r \sim R_{500}$), 
the deviation converges to similar amounts due to the limited impact of
cooling and feedback over large volumes: variations between $+0.03$
(weak winds) and $+0.10$ (strong winds) become comparable to $\Delta
\Omega_{\rm m}/\Omega_{\rm m} \approx +0.08$ as measured in
non--radiative runs.

A further contribution to the uncertainties comes from the dependence
upon the radius and redshift of the ratio $f_{\rm star}/f_{\rm gas}$.
In the observational determination of the baryon fraction from
eqn.~\ref{eq:cosmo}, this quantity is generally assumed to be $0.16$,
as measured in the Coma cluster within the virial radius [e.g. White
et al. 1993; note that our estimate from Lin et al. (2003) and
adopting the total mass measurements in Arnaud et al. (2005) is $0.11
\pm 0.04$].  If we compare the ratio $\phi = f_{\rm star}/f_{\rm gas}$
measured for a Coma--like simulated cluster at $R_{200}$ and $z=0$ 
with the estimates at other redshifts ($z_o =0.7$ and $1$; 
see, e.g., Fig.~\ref{fig:fgas_r}), we evaluate from
equation~\ref{eq:cosmo} 
\begin{equation}
\frac{\Omega_{\rm m}^{\prime} - \Omega_{\rm m}}{\Omega_{\rm m}} =
\frac{\Delta \Omega_{\rm m}}{\Omega_{\rm m}} =
\frac{ 1 + \phi(<R_{\Delta}, z=0) }{1 + \phi(<R_{\Delta}, z=z_o)} 
\; -1 \approx -0.05\,.
\end{equation}
When $\phi = f_{\rm star}/f_{\rm gas}$ measured locally at $R_{2500}$ is
compared with the corresponding value at different redshifts, the
deviation ranges between $-0.74$ (when winds are excluded) and $-0.38$
(when strong winds are present), whereas it is about $-0.10$ at
$R_{500}$.

As for the runs with gravitational heating only the effect of the
variation of $Y_{\rm b}$ with redshift and overdensity implies $\Delta
\Omega_{\rm m}/\Omega_{\rm m} \la +0.11$, thus comparable to the
current statistical uncertainties from {\it Chandra} observations of
the massive clusters out to $z=1.3$ (Ettori et al. 2003, Allen et
al. 2004). However, when the extra-physics of the radiative runs is
included, $\Delta \Omega_{\rm m}/\Omega_{\rm m}$ has two contributions
of $\approx +0.10$ and $\la -0.05$, due to an increase with redshift
of (1) $Y_{\rm b}$ (see Fig.~\ref{fig:ybar}) and (2) the stellar to
gas mass fraction ratio (see Fig.~\ref{fig:fgas_r}).  Both these
effects are caused by a more efficient star formation in high redshift
clusters.

In general, our results indicate that it may be dangerous to use
simulations to calibrate observational biases for precision
determination of cosmological parameters from the gas fraction in
clusters. Although none of our simulation models includes a fair
description of the actual ICM physics, it is interesting that
different models provide different redshift--dependent corrections for
the estimate of the cosmic baryon fraction from observations of the
gas and star density distribution within clusters. If applied to
observational data, such corrections would induce sizeable differences
in the determination of the matter and dark energy density parameters.

\section{Summary and discussion}

We have analyzed the distribution of gas and stellar mass fraction in
simulated massive X-ray galaxy clusters as function of (i) the radius
expressed in the form of $R_{2500}$, $R_{500}$ and $R_{200}$, (ii) the
redshift at which the structure is identified, (iii) the physical
processes determining the evolution of the baryons in the cluster
potential wells (i.e. gravitational heating, radiative cooling, star
formation, conduction and galactic winds powered by supernova
explosions).

Our main results can be summarized as follows.
\begin{enumerate}

\item As for the cluster set extracted from a cosmological box, which
are simulated by including cooling, star formation and feedback with
weak (340 km s$^{-1}$) winds, we find at $R_{200}$ $Y_{\rm b}=0.93$,
$Y_{\rm gas}=0.74$ and $Y_{\rm star}=0.20$, with scatter around these
values of 2, 4 and 8 per cent, respectively.  These results are
virtually independent of the cluster mass over the range $M_{\rm vir}
\approx (0.5 - 13) \times 10^{14} h^{-1}M_{\odot}$.  The dispersion relative
to the mean value measured at $R_{2500}$ is a factor of about $3$
larger than $R_{200}$.

\item In the four massive ($M_{200} > 10^{15} h^{-1}M_{\odot}$)
galaxy clusters simulated with 6 different physical schemes, we find
that the cosmic value of the baryon fraction $\Omega_{\rm
b}/\Omega_{\rm m}$ is reached at about $3 \times R_{200}$.  The
gas fraction increases radially, reaching 50 (80) per cent of the
value measured at $R_{200}$ at $r \approx 0.1 (0.3) R_{200}$
at $z=0$. At $z=1$ the same values are reached at radii which
are about 40 per cent larger. This indicates that $f_{\rm gas}$ tends
to be less concentrated at higher redshift, where the more efficient
star formation causes a more efficient removal of gas from the hot
phase in the central cluster regions. 
We also find that in these clusters the amount of
hot baryons, in unit of the cosmic value, is less scattered and
less dependent on the particular physics adopted when it is measured
over larger cluster regions. 
In the runs with {\it Gravitational heating only}, $Y_{\rm b} = f_{\rm bar} /
(\Omega_{\rm b}/\Omega_{\rm m})$ ranges between $0.82 \pm 0.06$ (at
$R_{2500}$) and $0.89 \pm 0.02$ (at $R_{200}$) at $z=0$. It increases
at $z=1$ to $0.91 \pm 0.10$, $0.95 \pm 0.04$ and
$0.94 \pm 0.03$ at $R_{2500}$, $R_{500}$ and $R_{200}$, respectively.
This increase of $Y_{\rm b}$ with $z$ owes to the smaller radii 
(in units of the virial radius)
at higher redshift where shock-heated baryons accrete, permitting 
the $f_{\rm b}$ values to reach the cosmic value at higher overdensities,
for example, at $z=1$ than at $z=0$ (see, e.g., Fig.~\ref{fig:yb_r}).
These values are roughly in agreement with observational
constraints obtained from highly X-ray luminous clusters (e.g. Ettori
\& Fabian 1999, Allen et al. 2004) at $R_{2500}$ and $R_{500}$.  
Using an SPH scheme with reduced viscosity, the non--thermal pressure
support contributed by turbulent gas motions, reduces the gas fraction
by 15 per cent at $R_{2500}$ and by $\sim$5 per cent at $r>R_{500}$.

Adding extra-physics to the action of gravity reduces the amount of
diffuse baryons by 20--40 (20--65) per cent at $R_{500}$ ($R_{2500}$).
This is due to the high star formation efficiency that is not
sufficiently mitigated by our assumed supernova feedback. The presence
of strong winds (assumed here $v \la 480$ km s$^{-1}$) induces the
following effects: (1) a reduction of $Y_{\rm star}$ by about a factor
of $2$ with respect to the case with no winds; (2) a milder radial
dependence of $Y_{\rm star}$ and $Y_{\rm gas}$; (3) larger values for
the ratio between $f_{\rm gas}$ and $f_{\rm star}$, that is however
still smaller, by about a factor $2.5$, than the observed values (Lin
et al. 2003) .

\item At $R_{2500}$, the estimate of $Y_{\rm b}$ is strongly affected
by the physics included in the simulations.  Super-cosmic values are
found for the radiative runs, as a consequence of the exceedingly
efficient cooling process, which causes a rapid sinking of baryons in
the cluster cores.  Moreover, differences in the formation history of
individual clusters contribute to a wider spread in the $Y_{\rm b}$
values, as shown in the runs including only gravitational heating. In
these cases, the standard deviation is of about 10 per cent.

\item At $R_{500}$ and $R_{200}$, the dispersion in the estimate of
$Y_{\rm b}$ is 2-3\% of the measured mean.  With respect to the mean
values measured at $R_{2500}$, $Y_{\rm star}$ decreases by a factor
$\ga 2$ and $Y_{\rm gas}$ increases by more than $1.3$, with peaks of
2-3 in the simulations without winds, or with weak winds combined with
conduction at $1/3$ the Spitzer value.

We explain the reduction in the scatter at $r>R_{500}$ by the common
history that the baryons experience when their properties are averaged
over a large enough volume. In the outer regions, the gas evolution is
mainly driven by the action of gravity. The
ratio $f_{\rm gas}/f_{\rm star}$, that is $\la 1$ at $R_{2500}$ and $z
\ga 0.3$, increases with radius and redshift and becomes $\simeq 3$
at $R_{200}$ and $z=0$ in the presence of winds.

\item Our results have direct implications on the calibration of the
systematic effects that limit the use of the X-ray gas mass fraction
as a cosmological tool (e.g. Evrard 1997, Ettori et al. 2003, Allen et
al. 2004). Owing to the uncertainties in the modelization of the ICM,
we emphasize that our simulations cannot be used to calibrate the
observed $f_{\rm gas}$. However, our results emphasize the potential
problems related to the estimate of the cosmic baryon fraction from
the gas mass fraction as measured in central cluster regions; i.e. the
variation of $Y_{\rm b}$ with redshift and a proper estimate 
of the contribution of stars in galaxies to the total cluster baryon 
budget as a function of radius and redshift.

For the non--radiative runs the effect of the variation of $Y_{\rm b}$
with redshift and overdensity implies a $\Delta \Omega_{\rm
m}/\Omega_{\rm m} \la +0.11$ that is comparable to the typical
statistical uncertainties from {\it Chandra} observations of the
massive clusters out to redshift $z=1.3$ (Ettori et al. 2003, Allen et
al. 2004). However, when star formation and feedback are also
included, $\Delta \Omega_{\rm m}/\Omega_{\rm m}$ has two contributions
of $\approx +0.10$ and $\la -0.05$, due to an increase with redshift
of (1) $Y_{\rm b}$ (see Fig.~\ref{fig:ybar}) and (2) ration between
stellar and gas mass (see Fig.~\ref{fig:fgas_r}).
\end{enumerate}

Our results show that star formation in our radiative runs is still
too efficient, especially in the cluster central regions, even in the
presence of rather strong galactic winds.  As a result, the impact of
cooling and star formation in the simulated clusters is still much
larger than in realistic cases.  On the one hand, this calls for the
need to introduce more effective feedback mechanisms, e.g. related to
Active Galactic Nuclei, which are able to prevent overcooling in the
central cluster regions (see, e.g., the effect of the injection 
of supersonic AGN jets in the ICM of simulated clusters in Zanni et al. 2005).
On the other hand, our analysis demonstrates
that using simulations to calibrate the systematic uncertainties in
the estimate of cosmological parameters from the cluster gas mass
fraction may be quite problematic, especially if it has to be used as
a high--precision tool to measure the cosmic density parameters
associated to matter and dark energy. 


\section*{ACKNOWLEDGEMENTS} 
The simulations have been performed using the IBM-SP4 machine at the
``Consorzio Interuniversitario del Nord-Est per il Calcolo
Elettronico'' (CINECA, Bologna), with CPU time assigned thanks to the
INAF--CINECA grant, the IBM-SP3 machine at the Italian Centre of
Excellence ``Science and Applications of Advanced Computational
Paradigms'', Padova and the IBM-SP4 machine at the ``Rechenzentrum der
Max-Planck-Gesellschaft'' at the ``Max-Planck-Institut f\"ur
Plasmaphysik'' with CPU time assigned to the ``Max-Planck-Institut
f\"ur Astrophysik''.  This work has been partially supported by the
INFN--PD51 grant and by MIUR. The anonymous referee is thanked 
for insightful comments.

\appendix
\begin{table*}
\caption{Distribution of the baryons in the two sets of
simulated galaxy clusters. The mean values of the indicated ratios
are shown.} \begin{tabular}{ 
 c@{\hspace{.6em}} c@{\hspace{.6em}} c@{\hspace{.6em}} c@{\hspace{.6em}}
 c@{\hspace{.6em}} c@{\hspace{.6em}} c@{\hspace{.6em}} c@{\hspace{.6em}}
 c@{\hspace{.6em}} c@{\hspace{.6em}} c@{\hspace{.6em}} c@{\hspace{.6em}}
 c@{\hspace{.6em}} c@{\hspace{.6em}} c@{\hspace{.6em}} }
\hline \\ 
STATUS & N & $z$ & & \multicolumn{3}{c}{$R_{2500}$} & & 
 \multicolumn{3}{c}{$R_{500}$} & & \multicolumn{3}{c}{$R_{200}$}  \\ 
 & & & & $Y_{\rm gas}$ & $Y_{\rm star}$ & $Y_{\rm b}$ & & 
 $Y_{\rm gas}$ & $Y_{\rm star}$ & $Y_{\rm b}$ & & 
 $Y_{\rm gas}$ & $Y_{\rm star}$ & $Y_{\rm b}$ \\ 
 \\ 
G & 4 & 0.0 & & $0.817 (0.064)$ & $-$ & $0.817 (0.064)$ & & $0.874 (0.023)$ & $-$ & $0.874 (0.023)$ & & $0.894 (0.020)$ & $-$ & $0.894 (0.020)$ \\ 
 ... & & 0.3 & & $0.864 (0.057)$ & $-$ & $0.864 (0.057)$ & & $0.907 (0.022)$ & $-$ & $0.907 (0.022)$ & & $0.926 (0.019)$ & $-$ & $0.926 (0.019)$ \\ 
 ... & & 0.7 & & $0.888 (0.020)$ & $-$ & $0.888 (0.020)$ & & $0.945 (0.028)$ & $-$ & $0.945 (0.028)$ & & $0.933 (0.027)$ & $-$ & $0.933 (0.027)$ \\ 
 ... & & 1.0 & & $0.910 (0.094)$ & $-$ & $0.910 (0.094)$ & & $0.947 (0.037)$ & $-$ & $0.947 (0.037)$ & & $0.942 (0.033)$ & $-$ & $0.942 (0.033)$ \\ 
 \\ 
GV & 4 & 0.0 & & $0.690 (0.045)$ & $-$ & $0.690 (0.045)$ & & $0.818 (0.018)$ & $-$ & $0.818 (0.018)$ & & $0.864 (0.010)$ & $-$ & $0.864 (0.010)$ \\ 
 ... & & 0.3 & & $0.720 (0.069)$ & $-$ & $0.720 (0.069)$ & & $0.865 (0.019)$ & $-$ & $0.865 (0.019)$ & & $0.900 (0.013)$ & $-$ & $0.900 (0.013)$ \\ 
 ... & & 0.7 & & $0.720 (0.033)$ & $-$ & $0.720 (0.033)$ & & $0.914 (0.038)$ & $-$ & $0.914 (0.038)$ & & $0.913 (0.025)$ & $-$ & $0.913 (0.025)$ \\ 
 ... & & 1.0 & & $0.722 (0.106)$ & $-$ & $0.722 (0.106)$ & & $0.912 (0.045)$ & $-$ & $0.912 (0.045)$ & & $0.935 (0.016)$ & $-$ & $0.935 (0.016)$ \\ 
 \\ 
FwW & 4 & 0.0 & & $0.513 (0.021)$ & $0.419 (0.029)$ &  $0.931 (0.021)$ & & $0.658 (0.029)$ & $0.262 (0.011)$ & $0.920 (0.026)$ & & $0.701 (0.027)$ & $0.226 (0.009)$ & $0.927 (0.018)$ \\ 
 ... & & 0.3 & & $0.448 (0.040)$ & $0.565 (0.054)$ & $1.013 (0.063)$ & & $0.664 (0.025)$ & $0.286 (0.022)$ & $0.950 (0.027)$ & & $0.708 (0.018)$ & $0.243 (0.018)$ & $0.951 (0.019)$ \\ 
 ... & & 0.7 & & $0.341 (0.059)$ & $0.749 (0.114)$ & $1.090 (0.060)$ & & $0.649 (0.024)$ & $0.305 (0.021)$ & $0.953 (0.014)$ & & $0.687 (0.028)$ & $0.259 (0.013)$ & $0.946 (0.019)$ \\ 
 ... & & 1.0 & & $0.288 (0.036)$ & $0.884 (0.098)$ & $1.172 (0.090)$ & & $0.631 (0.036)$ & $0.330 (0.038)$ & $0.961 (0.028)$ & & $0.684 (0.029)$ & $0.269 (0.019)$ & $0.953 (0.033)$ \\ 
 \\ 
FwWC & 4 & 0.0 & & $0.538 (0.058)$ & $0.398 (0.049)$ &  $0.935 (0.021)$ & & $0.655 (0.031)$ & $0.257 (0.015)$ & $0.912 (0.027)$ & & $0.714 (0.027)$ & $0.221 (0.011)$ & $0.936 (0.017)$ \\ 
 ... & & 0.3 & & $0.448 (0.070)$ & $0.552 (0.072)$ & $1.000 (0.063)$ & & $0.652 (0.026)$ & $0.281 (0.017)$ & $0.932 (0.025)$ & & $0.726 (0.026)$ & $0.236 (0.017)$ & $0.962 (0.014)$ \\ 
 ... & & 0.7 & & $0.302 (0.090)$ & $0.803 (0.169)$ & $1.105 (0.079)$ & & $0.631 (0.028)$ & $0.306 (0.018)$ & $0.937 (0.017)$ & & $0.707 (0.034)$ & $0.250 (0.008)$ & $0.958 (0.027)$ \\ 
 ... & & 1.0 & & $0.247 (0.052)$ & $0.965 (0.160)$ & $1.212 (0.126)$ & & $0.620 (0.058)$ & $0.327 (0.048)$ & $0.946 (0.025)$ & & $0.704 (0.033)$ & $0.263 (0.021)$ & $0.967 (0.027)$ \\ 
 \\ 
F & 1 & 0.0 & & $0.294 (-)$ & $0.682 (-)$ &  $0.976 (-)$ & & $0.535 (-)$ & $0.370 (-)$ & $0.905 (-)$ & & $0.593 (-)$ & $0.333 (-)$ & $0.927 (-)$ \\ 
 ... & & 0.3 & & $0.169 (-)$ & $1.083 (-)$ & $1.252 (-)$ & & $0.491 (-)$ & $0.447 (-)$ & $0.938 (-)$ & & $0.594 (-)$ & $0.369 (-)$ & $0.963 (-)$ \\ 
 ... & & 0.7 & & $0.123 (-)$ & $1.465 (-)$ & $1.588 (-)$ & & $0.456 (-)$ & $0.486 (-)$ & $0.942 (-)$ & & $0.564 (-)$ & $0.371 (-)$ & $0.935 (-)$ \\ 
 ... & & 1.0 & & $0.180 (-)$ & $1.054 (-)$ & $1.234 (-)$ & & $0.558 (-)$ & $0.409 (-)$ & $0.967 (-)$ & & $0.578 (-)$ & $0.362 (-)$ & $0.940 (-)$ \\ 
 \\ 
FsW & 1 & 0.0 & & $0.638 (-)$ & $0.302 (-)$ &  $0.939 (-)$ & & $0.699 (-)$ & $0.199 (-)$ & $0.898 (-)$ & & $0.747 (-)$ & $0.176 (-)$ & $0.923 (-)$ \\ 
 ... & & 0.3 & & $0.600 (-)$ & $0.439 (-)$ & $1.038 (-)$ & & $0.731 (-)$ & $0.245 (-)$ & $0.976 (-)$ & & $0.774 (-)$ & $0.203 (-)$ & $0.978 (-)$ \\ 
 ... & & 0.7 & & $0.439 (-)$ & $0.617 (-)$ & $1.055 (-)$ & & $0.718 (-)$ & $0.263 (-)$ & $0.981 (-)$ & & $0.740 (-)$ & $0.210 (-)$ & $0.951 (-)$ \\ 
 ... & & 1.0 & & $0.409 (-)$ & $0.552 (-)$ & $0.960 (-)$ & & $0.745 (-)$ & $0.248 (-)$ & $0.993 (-)$ & & $0.730 (-)$ & $0.207 (-)$ & $0.937 (-)$ \\ 
\hline \\ 
\end{tabular} 

\label{tab:fgas}
\end{table*}

\end{document}